\documentclass[aip,reprint,english,a4paper] {revtex4-1}

\usepackage{upgreek}
\usepackage{amsmath}
\usepackage{amsfonts}
\usepackage{amstext}
\usepackage{mathtools}
\usepackage{amssymb}
\usepackage{siunitx}
\usepackage[usenames]{color}
\usepackage[normalem]{ulem}
\usepackage{float}
\usepackage{tikz}

\newcommand{\dirfig}{.}
\newcommand*{\e}{\mathrm e}
\newcommand*{\mean}[1]{\left< #1 \right>}
\newcommand*{\up}[1]{\textsuperscript{#1}}

\usepackage{mathtools}
\newcommand{\itGamma}{{\it{\Gamma}}} 
\newcommand{\euler}{\mathrm{e}}
\newcommand{\kb}{\mathrm{k}_\mathrm{B}}
\newcommand{\EV}[1]{\left< #1 \right>}
\renewcommand{\vec}[1]{{\boldsymbol{#1}}}
\renewcommand{\d}{\mathrm{d}}
\DeclareSymbolFont{rmlargesymbols}{OMX}{mdbch}{m}{n}
\DeclareMathSymbol{\rmintop}{\mathop}{rmlargesymbols}{82}
\newcommand{\upint}{\rmintop\nolimits}
\newcommand{\im}{\mathrm{i}}

\newcommand{\SItwoDGammaVar}  {S1}
\newcommand{\SItwoDTI} {S2}
\newcommand{\SItwoDGammaUnit}  {S3}
\newcommand{\SINormal}  {S4}
\newcommand{\SITrypPCs}  {S5}

\begin{document}

\title{Path separation of dissipation-corrected targeted
molecular dynamics simulations of protein-ligand unbinding}

\author{Steffen Wolf$^{\dagger}$}%
\affiliation{Biomolecular Dynamics, Institute of Physics, Albert Ludwigs University, 79104 Freiburg, Germany}%
\email{steffen.wolf@physik.uni-freiburg.de;\\ stock@physik.uni-freiburg.de}
\altaffiliation{$^{\dagger}$These authors contributed equally to this work}
\author{Matthias Post$^{\dagger}$}
\affiliation{Biomolecular Dynamics, Institute of Physics, Albert Ludwigs University, 79104 Freiburg, Germany}%
\author{Gerhard Stock}%
\affiliation{Biomolecular Dynamics, Institute of Physics, Albert Ludwigs University, 79104 Freiburg, Germany}%
\email{stock@physik.uni-freiburg.de}
\date{\today}%

\begin{abstract}
Protein-ligand (un)binding simulations are a recent focus of biased 
molecular dynamics simulations. Such binding and unbinding can
occur via different pathways in and out of a binding site. We here
present a theoretical framework how to compute kinetics
along separate paths and to combine the path-specific rates into
global binding and unbinding rates for comparison with experiment.
Using dissipation-corrected targeted molecular dynamics
in combination with temperature-boosted Langevin equation simulations 
[Nat. Commun. \textbf{11}, 2918 (2020)]
applied to a two-dimensional model and the trypsin-benzamidine
complex as test systems, we assess the robustness of the procedure
and discuss aspects of its practical applicability to predict multisecond 
kinetics of complex biomolecular systems.
\end{abstract}

\maketitle
%
%
\section{Introduction}
\vspace*{-4mm}

Reaction paths (or pathways) constitute a central concept to describe
biophysical processes such as protein folding, allostery, and the
binding and unbinding of biomolecular complexes.\cite{Chipot07} When
we consider the free energy landscape $\Delta G(\vec{x})$ along a (in
general multidimensional) reaction coordinate $\vec{x}$ describing
such a process, the minima correspond to metastable states reflecting
the stationary distribution, while the valleys and saddle points
connecting the minima account for the flux and thus the transfer rates
between the states.\cite{Wales03} The connection between two
minima is in general not unique, i.e., various folding or unbinding
pathways can exist. Hence biophysical processes can be described via
path integrals,\cite{Faccioli06} sampling schemes such as
transition path sampling,\cite{Dellago16} or weighed
ensemble methods.\cite{Zuckerman17} The knowledge of reaction paths is
also useful to construct biasing coordinates, along which the sampling
of rare transition events can be enhanced.
\cite{Capelli19,Capelli19a,Rydzewski19,Bianciotto21,NunesAlves21,Henin22}

In this work, we further exploit the concept of reaction paths to facilitate the
general applicability of dissipation-corrected targeted molecular
dynamics (dcTMD),\cite{Wolf18,Wolf20,Post22} a recently developed
method to calculate slow (say, second-long) biomolecular kinetics from
atomistic simulations.\cite{Tiwary15,Plattner15,Teo16}
Combining enhanced sampling techniques with Langevin modeling, dcTMD
applies a constant velocity constraint to drive an atomistic system
from an initial state into a target state, say, the bound and unbound
states of a protein-ligand complex. To extract the equilibrium free
energy from these nonequilibrium simulations, dcTMD exploits
Jarzynski's identity\cite{Jarzynski97,Hendrix01,Dellago14}
\begin{align} \label{eq:JI} 
	\beta \Delta G = - \ln \EV{\euler^{-\beta W}},
\end{align}
which estimates the free energy difference $\Delta G$ between the two
states from the amount of work $W$ done on the system to
enforce the nonequilibrium process. Here $\EV{\ldots}$ denotes an
ensemble average over statistically independent nonequilibrium
simulations starting from a common equilibrium state, and
$\beta = 1/\kb T$ is the inverse temperature. To avoid problems
associated with the poor convergence behavior of the exponential
average,\cite{Ytreberg04,Echeverria12} we may exploit the second-order
cumulant expansion of Jarzynski's identity,
\begin{align} \label{eq:cumulant} 
	\Delta G \approx \EV{W} - \frac{\beta}{2} \EV{\delta W^2} ,
\end{align}
where the first term represents the averaged external work, while the
second term (with $\delta W = W-\EV{W}$) corresponds to the mean
dissipated work $\EV{W_{\rm diss}}$ of the process (and explains the
name 'dissipation-corrected' TMD). Since the knowledge of $W_{\rm diss}$
allows us to calculate the
friction coefficient $\itGamma$, in a second step the equilibrium
dynamics of the process can be studied by propagating a Langevin equation,
which is based on the free energy $\Delta G$ and the friction
$\itGamma$ obtained from dcTMD.\cite{Wolf20}

If the distribution of the external work $W$ is well approximated by a
Gaussian, the cumulant approximation in Eq.\ \eqref{eq:cumulant} is
known to become exact. Since this approximation is indispensable for the
efficient treatment of large molecular systems, the assumption of a
normal work distribution is in fact the main (and formally only) condition
underlying dcTMD. We expect such a distribution in the limit of slow
pulling, where the response of the system is linear and the work is
given as sum of many independent contributions,
\cite{Hendrix01,Wolf18} as required by the central limit
theorem.\cite{Billingsley95} In practice, though, this consideration
is mainly of academic interest, because the computational effort is
inversely proportional to the pulling velocity. 
By extending the argument to independent contributions in
space, Post et al.\cite{Post22} argued that a Gaussian
work distribution is also found for fluids. On the other hand, the assumption 
was shown to break down in
the case of ligand-protein dissociation processes that provide several
exit pathways of the ligand\cite{Wolf20} or ion diffusion through
a membrane channel coupled to a conformational change of
the protein.\cite{Jager22}

\begin{figure*}
\centering
\includegraphics[width=0.9\textwidth]{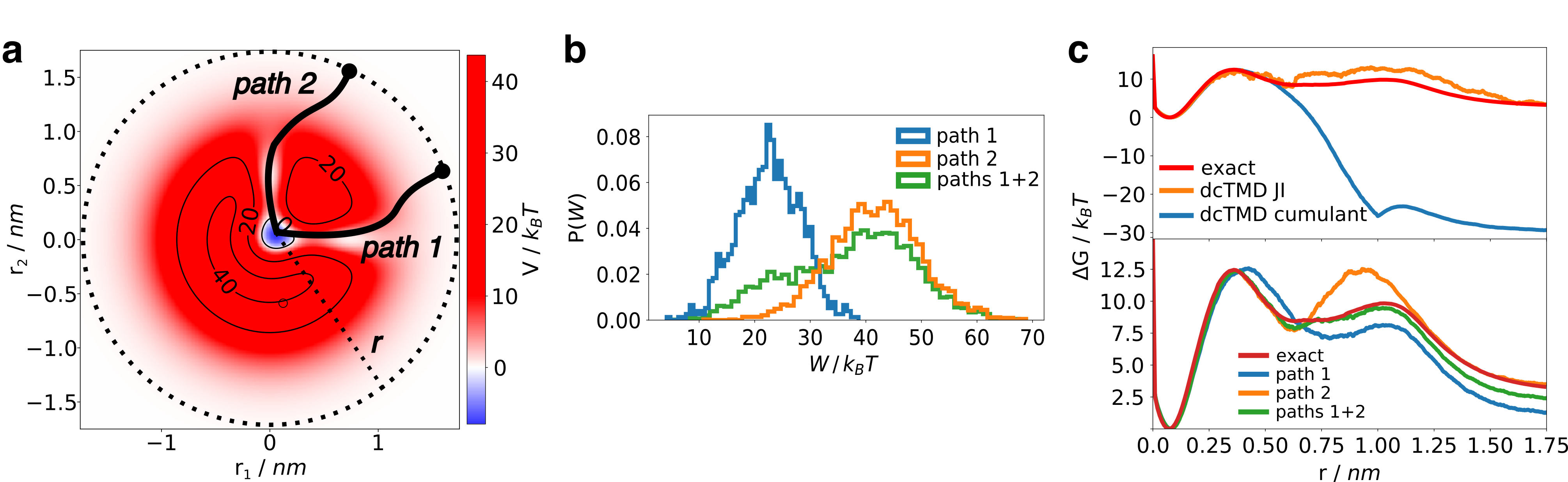}
\caption{
  2D model of ligand-protein dissociation,
   illustrating the separation of the total reaction flux into two
   paths 1 and 2. (a) Potential energy landscape $V(r,\theta)$, where
   the radial coordinate $r$ corresponds to the ligand-protein
   distance and the angle $\theta$ accounts for the direction of the
   process. (b) Total work distribution (green)
   compared to the distributions from the individual paths 1 (blue)
   and 2 (orange), obtained at $r=r_D$ from the dcTMD simulations. (c)
   Comparison of free energy profiles $\Delta G(r)$. (Top) Exact $\Delta G(r)$
   (red) as calculated from Eq.\ \eqref{eq:dG_proj}, compared to dcTMD
   simulations using Jarzynski's identity [Eq.\ \eqref{eq:JI}, in
   orange] and its cumulant approximation [Eq.\ \eqref{eq:cumulant},
   in blue], where we averaged over 5000 dcTMD runs. (Bottom)
   Free energy profiles for the individual paths 1 (blue) and 2
   (orange), and resulting dcTMD estimates of the total free energy
   (green) compared to the exact result (red).}
\label{fig:2DHam}
\end{figure*}

To explain the origin of this breakdown, we consider a minimal model
of ligand-protein dissociation, given by a two-dimensional (2D)
potential energy landscape $V(r,\theta)$ and corresponding friction 
profile $\itGamma(r,\theta)$ (see the Supplementary Material and 
Fig.~\SItwoDGammaVar\ for details). As shown in Fig.~\ref{fig:2DHam}a, 
the initial bound state is located at the origin ($r\!=\!0$), and the dissociated state is reached via external pulling along the radial coordinate $r$ to $r_D=1.75$~nm. While in most directions
(defined by the angle $\theta$), the ligand encounters high energy
barriers mimicking the steric hindrance of the surrounding protein,
two well-separated paths ('1' and '2') of relatively low energy exist,
along which dissociation may occur. In general, these paths will
exhibit different free energy curves as well as different friction
factors, but both end at the same energy, accounting for the
dissociated state.

Aiming to model the dissociation process via dcTMD, we pull 
the ligand along the radial coordinate $r$ accounting for
the ligand-protein distance, giving rise to the free energy profile
\begin{align} \label{eq:dG_proj}
\beta \Delta G(r) &= - \ln \left[ \upint_0^{\infty} \!\!r^\prime \,
              \d r^\prime  \delta(r-r^\prime) 
\, \upint_0^{2 \pi} \d \theta \; \e^{-\beta V(r^\prime, \theta) } \, \right]
              \nonumber \\ 
&= -  \ln \left[ \upint_0^{2 \pi} \!\!\d\theta \; \e^{-\beta
                              V(r, \theta) }  \right] - \ln r, 
\end{align}
where the factor $-\ln r$ represents the configurational entropy
arising from the Jacobian of the projection. \cite{Doudou09} As shown
in Fig.\ \ref{fig:2DHam}c, the exact free energy $\Delta G(r)$
calculated from Eq.\ \eqref{eq:dG_proj} exhibits (by design) two
maxima at $r\sim\,0.36$ and $1.03$~nm, before the ligand dissociates for large $r$. Performing dcTMD simulations in combination
with Jarzynski's identity \eqref{eq:JI}, the reference results for
$\Delta G(r)$ are only accurately reproduced for small $r$, but
deteriorate at larger distances due to sampling problems.
Invoking the cumulant approximation, on the other hand,
only the first barrier is reproduced, while the free energy goes to
nonphysically low values ($\sim -30\, \kb T$) for longer
ligand-protein distances. Figure \ref{fig:2DHam}b reveals the
reason for this drastic breakdown of the cumulant approximation by clearly
showing a non-Gaussian shape of the
corresponding work distribution, which leads to an overestimation of
the variance of the work and thus also of the dissipated work.

To present a remedy for this problem, we note that in our simple model
the dissociation can occur exclusively through either path 1 or
2. Hence we may consider trajectories following one of these paths
separately, and calculate the work distribution associated with this
path. To define a simple separatrix of the two paths, we use the
diagonal $r_1=r_2$ in Fig.~\ref{fig:2DHam}a for 0.44\,nm\,$\le r \le$
0.88\,nm. Interestingly, Fig.\ \ref{fig:2DHam}b reveals that the work
distributions of the individual paths are both well approximated by a
Gaussian, although of different means and widths (for an illustration of
the sampling per path, see Figs.~\SItwoDGammaVar d-f). Adding these
distributions (weighted by the probability that the corresponding path
is taken), we by design recover the total work distribution, which
readily explains its bimodal appearance. Most importantly, the finding
of normal work distributions for each path suggests that we may apply
the cumulant approximation for each path separately, in order to obtain
the correct free energy profiles for the two paths, as is shown in Fig.\
\ref{fig:2DHam}c. The theory of path separation including 
the calculation of observables from various paths is a central goal of 
this work and will be developed in Sec.~II D.

While the discussion above seems to concern a technical problem of
dcTMD, we note that it represents in fact a quite general issue. It
occurs whenever we perform enhanced sampling along a 1D biasing
coordinate, although the considered process in fact takes place on a
multidimensional energy landscape with several reaction channels. As
an illustration, Fig.\ \SItwoDTI\ shows various thermodynamic
integration protocols to calculate the free energy profile
$\Delta G(r)$ of the 2D model. Although thermodynamic integration is
in principle exact, it is found to typically fail in this
example. This is a consequence of incomplete sampling of the two paths
by the biasing due to the absence of jumps over the path-separating energy
barrier, which should also occur for other 1D biasing techniques such
as umbrella sampling.

Motivated by this simple example, in this work we present general
theoretical considerations under which conditions a separation of the
total reaction flux in several pathways is possible, in the sense that
it ensures the validity of the cumulant approximation and thus the
applicability of dcTMD. We also discuss the question which paths are
most important (i.e., exhibits the highest reaction rates), and which
paths can be safely neglected. In practical calculations, the
definition and detection of paths represent a nontrivial issue. To
this end, we consider nonequilibrium principal component analysis
\cite{Post19} as a generic dimensionality reduction approach, which
allows to determine and characterize various paths in a
low-dimensional space. Employing the trypsin-benzamidine complex
\cite{Guillain70,Marquart83} as a well-established model problem to
test enhanced sampling techniques,\cite{Buch11,Plattner15,Tiwary15,
  Teo16,Votapka17,Betz19} the performance and the potential of the
proposed path separation scheme is discussed.

%
%
\section{Theory}
\vspace*{-4mm}
\subsection{Dissipation-corrected targeted molecular
dynamics } \label{sec:dcTMD}
\vspace*{-4mm}

For further reference, we first briefly summarize the theoretical
foundation of dcTMD, which was given in Refs.\
\onlinecite{Wolf18,Wolf20,Post22}. To pull two molecules apart (e.g.,
protein and ligand), TMD as developed by Schlitter et al.
\cite{Schlitter94} uses a constraint force $f$ that results in a
moving distance constraint $x = x_0 + v t$ with a constant velocity
$v$. Pulling from $x_0$ to $x$, the mean external work performed on the
system is given by
\begin{equation} \label{eq:Work}
\EV{W(x)} = \upint_{x_0}^{x} \d x^\prime \EV{f(x^\prime)}  \, ,
\end{equation}
where $\EV{\ldots}$ again denotes the average over all nonequilibrium
trajectories. Evaluating the second cumulant of the work in Eq.\
\eqref{eq:cumulant}, we obtain the dissipated work\cite{Wolf18}
\begin{align} \label{eq:Wdiss}
\EV{W_{\rm diss}(x)} &= \frac{\beta}{2} \upint_{x_0}^{x} \! \d x^\prime
                  \upint_{x_0}^{x} \! \d x'' 
	\EV{\delta f(x^\prime) \delta f(x'')} \, ,
\end{align}
which is defined via the autocorrelation function of the constraint
force $\delta f = f - \EV{f}$.

Exploiting that the cumulant approximation neglects higher orders of
the force autocorrelation function and due the simplification arising from
the constraint $x = x_0 + v t$, we can --without further
approximation-- derive a generalized Langevin equation
of the nonequilibrium process,\cite{Post22}
\begin{equation}\label{eq:GLE}
	m \ddot{x}(t) = - \frac{\partial G}{\partial x} - \upint^t_0
        \! \d \tau K(t,\tau) \, \dot{x}(\tau)  + \eta(t)  + f(t) = 0 \, ,  
\end{equation}
which includes the mean force $- \partial G/\partial x$, a
non-Markovian friction force with memory function $K(t,\tau)$, a
stochastic force $\eta (t)$ due to the associated colored noise with
zero mean, and the external pulling force $f(t)$. The memory kernel is
given by the autocorrelation function of the constraint force,
\begin{equation}\label{eq:memory}
	K(t_2, t_1) = \beta \EV{ \delta f(t_2) \delta f(t_1) } \, .
\end{equation}
It obeys the fluctuation dissipation theorem
\begin{equation} \label{eq:FDT}
	 K(t_2,t_1) = \beta \EV{ \eta(t_2) \eta(t_1) },
\end{equation}
which states that non-white noise $\eta$ causes a finite decay time of
the memory kernel $K$.  
Nevertheless, by using $\dot{x}(\tau)=v$, Eq.~\eqref{eq:GLE} can be
written in the form of a Markovian Langevin equation
\begin{equation}\label{eq:LE}
	m \ddot{x}(t) = - \frac{\partial G}{\partial x} - \itGamma(x(t))\,
	\dot{x}(t) + \eta(t)  + f(t) = 0 ,
\end{equation}
although the noise $\eta (t)$ is not delta-correlated due to Eq.~\eqref{eq:FDT}.
Here
\begin{equation}\label{eq:Gamma}
\itGamma(x) = \upint_{0}^{t(x)} \d t^\prime K(t,t^\prime)
\end{equation}
represents the position-dependent friction $\itGamma(x)$ associated
with the pulling process. Employing Eq.\ \eqref{eq:Wdiss}, the
friction can also be directly calculated from the dissipated work
via\cite{Post22}
\begin{equation}\label{eq:Gamma2}
 \itGamma(x) = \frac{\beta}{2 \, v} \, \frac{\partial }{\partial  x}
 \mean{ \delta W^2(x)},
\end{equation}
which is numerically advantageous compared to Eq.\ (\ref{eq:Gamma}),
as it avoids the computationally expensive calculation of
autocorrelation functions.
Data analysis was performed using Python packages
	NumPy,\cite{Harris20} SciPy,\cite{Virtanen20} and
	Pandas.\cite{McKinney10} Matplotlib\cite{hunter07} and
	Gnuplot\cite{gnuplot} were used for generating plots.
	Calculations with the 2D model were carried out using 
	a Jupyter notebook.\cite{Kluyver16} 

%
%
\subsection{Calculation of reaction rates} \label{sec:rates}
\vspace*{-4mm}

The Langevin model may be used to discuss the nonequilibrium process
in terms of the memory function $K(t_2,t_1)$ and the associated
friction $\itGamma(x)$, which both are directly obtained from
dcTMD. What is more, we can employ the free energy $\Delta G(x)$ and
the friction $\itGamma(x)$ from dcTMD to simulate the {\em unbiased}
dynamics via the free Langevin equation [Eq.\ (\ref{eq:LE}) with
$f=0$]. While Jarzynski's identity [Eq.\ \eqref{eq:JI}] states that
this is correct for the free energy, the friction determined by dcTMD
in general depends on the pulling velocity $v$, see Ref.\
\onlinecite{Post22}. Since this velocity dependence affects the overall transition rates (typically by a factor of $\sim 2$) clearly less
than errors in the estimate of the energy barrier, the usage of dcTMD
friction factors in equilibrium Langevin simulations nevertheless
represents a promising strategy to estimate transition rates of slow
processes.\cite{Wolf20}

The theoretical formulation of dcTMD given above can be nicely
illustrated by a simple model proposed by Zwanzig,\cite{Zwanzig73}
which couples a general 1D system (in general nonlinearly) to a
harmonic bath. In the unbiased case (with constraint force $f=0$), it
is well established that the model gives rise to an exact general
Langevin equation as in Eq.\ \eqref{eq:LE}. As shown in the Appendix,
the formulation readily generalizes to the case of
constrained dynamics used in dcTMD, revealing that both biased and
unbiased dynamics can be described by the same Langevin model.
Moreover, we derive explicit expressions for the resulting external
force $f(t)$ and the associated work $W$. Starting from a Boltzmann
distribution (as required for Jarzynski's identity), in particular, we
obtain a Gaussian distribution of the work, which renders the cumulant
approximation exact.

When we intend to model very slow processes such as ligand-protein
dissociation that can take seconds to hours, there is one more
practical issue to overcome. That is, since the numerical integration
of the Langevin equation typically require time steps of a few
femtoseconds, we would need to propagate Eq.\ (\ref{eq:LE}) for
$ \gtrsim 100 \cdot 10^{15}$ steps to sufficiently sample a process
occurring on a timescale of seconds. As a remedy, we employ the method
of ``$T$-boosting'' which
exploits the fact that temperature is the driving force of the
Langevin dynamics.\cite{Wolf20} That is, when we consider a process described by a
transition rate $k$ and increase the temperature from $T_1$ to $T_2$,
the corresponding rates $k_1$ and $k_2$ are related by the
Kramers-type expression
\begin{equation}\label{eq:Tboost}
k_2 = k_1 \euler^{- \Delta G^{\neq}(1/(\kb T_2) -1/(\kb T_1))} \, ,
\end{equation}
where $\Delta G^{\neq}$ denotes the transition state energy of the
process. Hence, by increasing the temperature we also increase the
number $n$ of observed transition events according to
$n_2/n_1 = k_2/k_1$. 
Unlike methods like temperature-accelerated MD,\cite{Sorensen00}
where $\Delta G(x)$ is first calculated at a high temperature and
subsequently rescaled to a desired low temperature (whereupon
$\Delta G(x)$ in general changes), $T$-boosting does not need to
assume that $\Delta G(x)$ is independent of $T$, because by using
dcTMD we calculate $\Delta G(x)$ right away at the desired
temperature. See Ref.~\onlinecite{Wolf20} for a discussion of
the performance of $T$-boosting for the trypsin-benzamidine complex.

%
%
\vspace*{-2mm}
\subsection{Validity of the cumulant approximation}
\vspace*{-4mm}

Since the main presumption of dcTMD is that the distribution of the
work $W$ resembles a Gaussian to justify the cumulant approximation
\eqref{eq:cumulant}, it is important to study the validity of this
assumption. To this end, we integrate Eq.\ \eqref{eq:LE} from the
initial position $x_0$ to the end position $x$, yielding
\begin{equation} \label{eq:singleTrajWork}
W(x) = \Delta G(x) + v \upint_{x_0}^{x} \!\! \d x^\prime
\itGamma(x^\prime) - \upint_{x_0}^{x} \!\! \d x^\prime \eta(x^\prime),
\end{equation}
where we defined $\eta(t)=\eta((x\!-\!x_0)/v)\equiv\eta(x)$. Upon
averaging, the stochastic force term cancels out (since
$\EV{\eta(x)}\!=\!0$), which again leads to the relation
$\Delta G = \EV{W} - \EV{W_{\rm diss}}$ discussed above. On the other
hand, when we consider the work $W$ associated with a single
trajectory, we find that the first two terms on the right-hand-side of
Eq.\ \eqref{eq:singleTrajWork} are identical for all
trajectories, due to the constraint $x=x_0+vt$. Hence the 
distribution of the work, $P(W)$, is
completely determined by the distribution $P(\delta W_{\rm diss})$ of
the integrated noise
$\delta W_{\rm diss}(x) \equiv \upint_{x_0}^{x} \!\! \d x^\prime
\eta(x^\prime)$. In other words, if $P(\delta W_{\rm diss})$ is normal
distributed, so is the work distribution $P(W)$.

Given as a sum of many small noise increments $\eta(x)$, the central
limit theorem generally assures that $P(\delta W_{\rm diss})$ is a
normal distribution. Importantly, this holds even if the noise $\eta$
arising in the Langevin equation is not Gaussian
distributed; only the integrated noise
$\delta W_{\rm diss}$ needs to be Gaussian. Moreover this conclusion
holds for position-dependent noise and friction, as long as the noise
integral is not dominated by some singularly high noise increments,
for which the central limit theorem breaks down. \cite{Billingsley95}
Given these conditions and the special case that the problem can
be properly described by a single reaction coordinate $x$, we
therefore expect a Gaussian work distribution and thus the validity of
the cumulant approximation.

The situation becomes considerably more involved, when the system
cannot be described by a single reaction coordinate $x$, but requires
another (in general multidimensional) coordinate $\vec{y}$ for an
appropriate description. Assuming for simplicity a 2D model, this
means that the system is pulled along coordinate $x$, but additionally
can freely evolve along coordinate $y$. Potential implications
of the additional degree of freedom on the resulting work distribution
have been illustrated in Fig.\ \ref{fig:2DHam} for a simple model
problem, where the motion along $y=\theta$ gives rise to a splitting of the
2D energy landscape into two 1D pathways (See also Fig.~\SItwoDGammaUnit). While the work distributions
of the individual paths are well approximated by a Gaussian, their
means and widths are different, such that the sum of the Gaussians
exhibits a bimodal appearance. As a consequence, we cannot simply
apply the cumulant approximation to the full problem, but need to
resort to a treatment of the individual paths.

We note in passing that in practice it is not obvious how to test the
Gaussianity of the work distribution, in order to ensure that the
cumulant approximation works. While well-established normality tests
exist, \cite{Henderson06} we found their results rather inconclusive
for the problem at hand, especially for the small number of
trajectories that are available for atomistic protein-ligand models.
As a qualitative means to study the shape of the work distribution
$P(W)$, especially its tail for low $W$, we may compare $P(W)$ to a
normal distribution whose mean and variance matches the sampled mean
and variance. In practice, we compare the points of the computed work
to the quantiles of the corresponding normal distribution, such that
the deviations reveal a non-Gaussian shape. This has the advantage
that the comparison explicitly shows all data points and does not
depend on the choice of bins. The test is demonstrated in Fig.\ \SINormal\
for the 2D model, which exhibits deviations from normality
in the full trajectory set that become resolved in the path-separated
sets.

%
%
\subsection{Path separation} \label{sec:PathSep}
\vspace*{-4mm}

To study under which conditions a separation of the total reaction
flux into several non-interacting pathways is possible and helpful for
practical calculations, we now present a mathematical description of
the path separation scheme discussed above. To this end, we again
assume that the considered process is well represented as a function
of the pulling coordinate $x$ and of a further coordinate $\vec{y}$,
along which the system can evolve freely. Given the resulting free
energy landscape $\Delta G(x,\vec{y})$, we wish to define a reaction
path $k$ as a class of similar trajectories $\left( x, \vec{y}\right)(t)$,
whose distribution can be characterized by a 1D 
free energy curve $\Delta G_k(x)$, which
describes this specific path from an initial (e.g., bound)
state to a final (e.g., unbound) state. This definition corresponds
to the assumption of distinct ''valleys'' in $\Delta G(x,\vec{y})$
connecting $x_0$ and $x_{\rm end}$. For simplicity, we for now assume
that trajectories of different paths should be mutually exclusive,
i.e., the paths should not overlap. As the latter condition may be
difficult to satisfy in practice, we will discuss in Sec.\
\ref{sec:praxis} a procedure to relax this condition.

Given the free energy landscapes $\Delta G(x,\vec{y})$ and
$\Delta G(x)$, we first introduce the probabilities
\begin{align}  
  P(x,\vec{y}) &= \euler^{-\beta \Delta G(x,\vec{y})}/Z
\label{eq:Pxy}, \\
  P(x) &= \euler^{-\beta \Delta G(x)}/Z
         = \upint \!\! \d \vec{y} \, \euler^{-\beta \Delta G(x,\vec{y})}/Z
\label{eq:Px}, \\
  Z &= \upint \!\! \d x\, \euler^{-\beta \Delta G(x)}
      = \upint \!\! \d x \! \upint \!\! \d \vec{y} \, \euler^{-\beta
      \Delta G(x,\vec{y})} 
\label{eq:Z}.
\end{align}
To partition the probability $P(x)$ into paths labeled by
$k \!=\! 1,\ldots,K$, we define regions $\vec{Y}_k(x)$
of the free energy
landscape $\Delta G(x,\vec{y})$, such that they only include
trajectories of path $k$, i.e.,
$\vec{Y}_k(x) = \{ \vec{y} \, | \, (x,\vec{y}) \in \text{path}\, k \}$. This allows
us to write
\begin{align} \label{eq:pathsum}
P(x) &= \sum_k P(x,k) , \\
 P(x,k) &= \frac{1}{Z} \upint_{\vec{y}\in\vec{Y}_k(x)}
\!\!\!\!\!\!\!\!\!\!\!\!\!\!\! \d \vec{y}\,
 \mathrm{e}^{-\beta \Delta G(x,\vec{y})}  ,
\end{align}
where $P(x,k)$ represents the joint probability to be on path $k$ and at
position $x$. Additional regions of the free energy landscape
$\Delta G(x,\vec{y})$ not covered by the reaction paths (such as
hardly accessible high-energy regions) can formally be included via
additional terms of the sum, but are not of interest here, as they
have a negligible impact on the integral in Eq.~\eqref{eq:pathsum}.
Alternatively, we may write
\begin{align} \label{eq:pathsum2} 
  P(x,k) = p_k P(x|k) ,
\end{align}
where $p_k$ is the probability to be on path $k$ and $P(x|k)$ is the resulting 
conditional probability to be at position $x$ if path $k$ is followed. 
In this way, specified by adding the subscript or	superscript ``eq'', we may 
interpret the partitioning in Eqs.\ \eqref{eq:pathsum} and
\eqref{eq:pathsum2} in terms of the free energy as
\begin{align}  \label{eq:pathsum3} 
P_{\rm eq}(x,k) \equiv  p_k^{\rm eq} \, \euler^{-\beta \Delta G_k(x)}/Z,
\end{align}
which yields the equilibrium path probabilities $p_k^{\rm eq}$ and path 
free energy curves $\Delta G_k(x)$. Both quantities are needed when we
want to calculate the reaction rates of the system via a Langevin
model as will be described in Sec.\ \ref{sec:pathImp}. 

Since we typically cannot calculate these quantities directly 
(due to the MD simulation time limits mentioned before), 
we perform nonequilibrium dcTMD simulations. 
Let us assume that we have in total $N$ nonequilibrium trajectories
labeled by $n \!=\! 1,\ldots,N$, which are divided into the $K$
different paths, each containing $N_k$
trajectories labeled by $n_k\!=\! 1,\ldots,N_k$. Due to the initial
preparation of the system and the nature of the paths, each path
occurs in the dcTMD simulations with an observed
probability $p_k^{\rm neq} = N_k/N$.
Expressing the ensemble average $\EV{\ldots}$ as a sum over all
nonequilibrium trajectories $n$ with work $W_n(x)$, we obtain from
Jarzynski's identity [Eq.~\eqref{eq:JI}]
\begin{align} \label{eq:pathsum4a} 
\euler^{-\beta \Delta G(x)} &= \frac{1}{N} \sum_{n=1}^N \euler^{-\beta W_n(x)}
  = \sum_{k=1}^K  \frac{N_k}{N} \frac{1}{N_k} \! \sum_{n_{k}=1}^{N_k}
    \euler^{-\beta W_{n_k}\! (x)}
  \nonumber \\
  & = \sum_{k} p_k^{\rm neq} \EV{\euler^{-\beta W(x)}}_k 
\equiv  \sum_{k} p_k^{\rm neq}  \, \euler^{-\beta \Delta \mathcal{G}_k(x)} \, ,
  \nonumber \\
  & \equiv Z \sum_k P_{\rm neq}(x,k),
\end{align}
where $\EV{\ldots}_k$ denotes the average over all trajectories of
path $k$, and $\Delta \mathcal{G}_k(x)$ represents the energy profile
along this path. Note that the latter can {\em per se} not be identified with
the free energy curves $\Delta G_k(x)$ in Eq.\ \eqref{eq:pathsum3},
because the path energy curves $\Delta \mathcal{G}_k(x)$ as
well as the path probabilities $p_k^{\rm neq}$ are nonequilibrium
quantities that depend on the protocol of the dcTMD simulations such
as the pulling velocity $v$.

Since Jarzynski's identity can be generalized\cite{Post19} to estimate
the free energy landscape $\Delta G(x,\vec{y})$, we may project the
corresponding joint probability $P(x,\vec{y})$ onto each path $k$
along coordinate $x$, yielding $P_{\rm neq}(x,k) = P_{\rm eq}(x,k)$.
This equivalence can be used to calculate the equilibrium path
probabilities $p_k^{\rm eq}$ from the nonequilibrium simulations via
\begin{align}  
  p_k^{\rm eq} &\equiv \upint \d x P_{\rm eq}(x,k)
  \label{eq:pathsum4b} \\
  &= \frac{p_k^{\rm neq}}{Z} \upint \d x\, \euler^{-\beta \Delta
    \mathcal{G}_k(x)} , \label{eq:pathsum4}
\end{align}
where the partition function
\begin{align}\label{eq:Zworking}
  Z =  \sum_{k} p_k^{\rm neq} \upint \!\! \d x \,  \euler^{-\beta
  \Delta \mathcal{G}_k(x)}
\end{align}
is obtained from the the path estimates in Eq.\ \eqref{eq:pathsum4a}.
(To derive Eq.\ \eqref{eq:pathsum4}, we used that 
$Z = \upint \d x\, \euler^{-\beta \Delta
    G(x)} = \upint \d x\, \euler^{-\beta \Delta
    G_k(x)}$ for any path $k$, which is a consequence of the normalization
  condition $\upint \d x\, P(x|k) = 1$.)
Moreover, the equivalence shows that the free energy curves
$\Delta G_k(x)$ in Eq.\ \eqref{eq:pathsum4a} and the corresponding
nonequilibrium energy curves $\Delta \mathcal{G}_k(x)$ in Eq.\
\eqref{eq:pathsum4} simply differ by a constant shift of the energy
\begin{equation} \label{eq:pathsum5} 
  \Delta G_k(x) = \Delta \mathcal{G}_k(x) + \beta^{-1}
  \ln \left( p^\text{eq}_k / p^\text{neq}_k \right) \, .
\end{equation}
As a consequence, both energy curves give rise to the same mean force
$f_{\rm PMF} = -\partial G / \partial x$ in Langevin equation
\eqref{eq:LE}, and can therefore both be considered as potentials of
mean force.

Just as for the general Jarzynski identity \eqref{eq:JI}, the
direct evaluation of the exponential average
$\EV{\euler^{-\beta W(x)}}_k$ of path $k$ in Eq.\ \eqref{eq:pathsum4a} is
numerically cumbersome. However, when we assume that the work
associated with each individual path is normal distributed, the 
potential of mean force
along path $k$ can be calculated using the
cumulant approximation,
\begin{align} \label{eq:Pathcumulant} 
\Delta \mathcal{G}_k(x) \approx \EV{W(x)}_k - \frac{\beta}{2}
  \EV{\delta W^2(x)}_k , 
\end{align}
which was the main motivation to perform the path separation in the
first place. 


%
%
\subsection{Path importance} \label{sec:pathImp}
\vspace*{-4mm}

When studying ligand-protein dissociation, a natural indicator for the
importance of a reaction path is the reaction rate
at equilibrium. This is because the total dissociation rate
$k_{\rm tot}$ is given as the sum over all equilibrium rates
contributing to the considered process. Assuming that path $k$ occurs in
an equilibrium simulation with probability $p^\text{eq}_k$, the total
rate can be written as
\begin{align} \label{eq:pathsum6} 
k_{\rm tot} = \sum_k p^\text{eq}_k k_k \, ,
\end{align}
where $k_k$ is the rate associated with the path. We note that
Eq.~\eqref{eq:pathsum6} corresponds to the well-known sum\cite{Logan96}
$k_{\rm tot} = \sum_k \tilde k_k$ of weighted first-order reaction
rates $\tilde k_k = p^\text{eq}_k k_k$. The weights
  $p^\text{eq}_k$ account for the correct partitioning of the process
  into paths, because the calculation of a rate along a single path
  does not include the possibility to switch between various paths.
Since a single path with an extraordinary high rate may completely
dominate the reaction, it is clearly of interest to identify the most
important pathways of a reaction. In this way, several paths of
similar importance may exist and therefore all contribute to
$k_{\rm tot}$.

As an illustrative example, we again consider the simple model
discussed in Fig.\ \ref{fig:2DHam}, which shows two reaction paths
($k=1,2$) with nonequilibrium energy curves $\Delta
\mathcal{G}_k(r)$. Using these curves and the associated path
probabilities $p_1^\text{neq} \approx 0.81$ and
$p_2^\text{neq} \approx 0.19$ from dcTMD, Fig.\ \ref{fig:2DHam}c shows
that the correct overall free energy $\Delta G(r)$ is indeed nicely
recovered from Eq.\ (\ref{eq:pathsum4a}). (As discussed in Sec.\
\ref{sec:praxis}, we exclude dcTMD trajectories that initially cross
between paths 1 and 2 (about 4\,\%) from the analysis.)  The
corresponding equilibrium path probabilities
$p_1^\text{eq} \approx 0.79$ and $p_2^\text{eq} \approx 0.21$
obtained from Eq.\ (\ref{eq:pathsum4}) are quite similar to the
nonequilibrium results, which indicates that $x$ is a suitable reaction coordinate.

To obtain the dissociation and association rates of the model, we
perform $10\times 1$~ms-long unbiased Langevin simulations [Eq.\
(\ref{eq:LE}) with $f\!=\!0$] independently for each 1D path (see the
Supplementary Material for details). As shown in Tab.~\ref{tab:rates}, the
two paths differ in their importance for both binding and unbinding.
While path\,2 exhibits larger rates for both processes, path\,1 is
accessed in 80~\% of all attempts, such that path\,1 has the largest
impact on the overall binding and unbinding rate.  Using Eq.\
(\ref{eq:pathsum6}), we also calculated the total dissociation and
association rates, i.e., averaged over the two paths. This is of
particular interest, because for the simple 2D model the total rate
can also be directly calculated by performing Langevin simulations on
the 'full-dimensional' potential energy landscape, i.e., without
performing a separation into 1D paths (see Fig.~\SItwoDGammaUnit\ for
details).  Table~\ref{tab:rates} shows perfect agreement of path-based
rates and 2D results for dissociation, while the association rates are
found to differ by 14\,\%.

\begin{table}[htbp]
   \centering
   \begin{tabular}{@{} lcccc @{}}
\hline
 & $p^\text{neq}_k$ & $p^\text{eq}_k$ & $k_{\rm off}$ (ms$^{-1}$) & $k_{\rm on}$ (ms$^{-1}$) \\
      \hline
      path\,1 \quad   & $0.81 \pm 0.01$  & $0.79 \pm 0.01$  & $16 \pm 2$ & $210 \pm 20$ \\
      path\,2 \quad  & $0.19 \pm 0.01$  & $0.21 \pm 0.01$ & $55 \pm 3$ & $230 \pm 10$ \\
      total \quad & -- & -- & $24 \pm 2$ & $220 \pm 10$ \\
      \hline
      total (2D) \quad   & -- & -- 		& $24 \pm 2$  & $190 \pm 10$ \\
      \hline
   \end{tabular}
   \caption{
     Dissociation ($k_{\rm off}$) and
     association ($k_{\rm on}$) rates of paths $k=1,\, 2$ of the 2D
     model in Fig.\ \ref{fig:2DHam}, obtained from unbiased Langevin
     simulations using the path free energy profiles
     $\mathcal{G}_k(x)$. Furthermore, nonequilibrium
     ($p^\text{neq}_k$) and equilibrium ($p^\text{eq}_k$) weights of
     the two paths are given, which are used to obtain the total rates
     via Eq.\ (\ref{eq:pathsum6}). For comparison, total rates are
     also calculated from Langevin simulations on the 2D potential
     energy landscape. Values in brackets denote the standard error of
     the mean. The error of the total rates
     from Eq.\ (\ref{eq:pathsum6}) is obtained from error
     propagation.}
   \label{tab:rates}
\end{table}

The above example demonstrates that even for a apparently simple model
problem, the interpretation of the various pathways contributing to
the reaction rate may be not straightforward. Rather than resorting to
approximate theories such as Kramers' relation (which, e.g., assumes a
single barrier of parabolic shape and constant friction), we therefore
recommend the calculation of reaction rates via the numerical
propagation of the unbiased Langevin equation.

%
%
\subsection{Path detection} \label{sec:pathDetection}
\vspace*{-4mm}

In our introductory example shown in Fig.~\ref{fig:2DHam}, the two
paths of the model are readily identified via visual inspection of the
2D potential energy landscape. When we consider an all-atom MD
simulation of a protein-ligand complex, however, this is typically not
possible because the trajectories evolve in a $3N\!-\!6$-dimensional
space of the $N$ atoms of the complex. Hence in a first step, we need
to employ some dimensionality reduction approach to identify
collective variables that appropriately describe the
unbinding dynamics in a low-dimensional space,\cite{Noe17,Sittel18}
and allow to cluster the unbinding trajectories into different
reaction pathways.

While various techniques to identify ligand
unbinding paths have been suggested,
\cite{Votapka17,Capelli19,Rydzewski19,NunesAlves21,Bianciotto21} 
here we employ a principal component analysis (PCA) of protein-ligand
contacts, \cite{Ernst15} which assumes that these contacts are
suitable features to represent the unbinding process.
Considering an equilibrium MD simulation with coordinates $\vec{q} =
\{q_i\}$, the basic idea is to construct the
covariance matrix
\begin{equation}\label{eq:CovEq}
\sigma_{ij}^{\rm eq} = \mean{ \delta q_i \delta q_j }_{\rm eq},
\end{equation}
where $\delta q_i \!=\! q_i \!-\! \mean{q_i}_{\rm eq}$. PCA
represents a linear transformation that diagonalizes $\sigma$ and thus
removes the instantaneous linear correlations among the
variables. Ordering the eigenvalues of eigenvectors
$\vec{e}_n^{\rm eq}$ decreasingly, the first principal components
$V_{n}^{\rm eq} = \vec{e}_n^{\rm eq} \cdot  \vec{\delta q}$ 
account for the directions of largest variance of the data, and are
therefore often used as collective variables.\cite{Amadei93,Altis08}

When we consider dcTMD simulations that enforce the unbinding of a
complex along the pulling coordinate $x$, we wish to
identify the directions of largest variance of the dcTMD data, in
order to separate the unbinding trajectories into different
paths. Hence we calculate the covariances directly from the dcTMD
simulations, and average over $x$ to obtain the nonequilibrium
covariance matrix\cite{Post19}
\begin{equation}\label{eq:CovAvNEQ}
\sigma^{\rm neq}_{ij} = \frac{1}{x_{\rm end} - x_0} \upint_{x_0}^{x_{\rm end}} \d x\,  
  \mean{ \delta q_i(x) \delta q_j(x) }_{{\rm neq}}.
\end{equation}
This leads to principal components $V^{\rm neq}_{n}$ that map out an
energy landscape $\Delta \mathcal{G}$ associated with the
nonequilibrium distribution generated by dcTMD. In practice, this
simply means to perform a PCA of the concatenated TMD trajectories.
Performing the PCA on contact distances, the first principal
  component by design correlates with the pulling coordinate. This
  connection may not be given if other input features, such as
  dihedral angles, are employed. In this case, we can consider the
  content of eigenvectors to identify their respective dynamics in
  Cartesian space.

Note that this nonequilibrium PCA is well defined, because the
nonequilibrium principal components are related to their equilibrium
counterparts via Jarzynski's identity. That is, when we replace the
equilibrium average in Eq.\ (\ref{eq:CovEq}) by a Jarzynski-type
average over the nonequilibrium data, we obtain\cite{Post19,Hummer01}
\begin{align}\label{eq:CovEqJ}
\sigma_{ij}^{\rm eq} &= \frac{\upint\!\! \d x \mean{\delta q_i(x) \delta q_j(x) \,
              \e^{-\beta W(x)} }_{\rm neq}}{\upint\!\! \d x^\prime
              \mean{ \e^{-\beta W(x^\prime)} }_{\rm neq}} \, ,
\end{align}
which is an equilibrium covariance matrix constructed via the
reweighting of the nonequilibrium data. We note that the
dimensionality reduction of nonequilibrium dcTMD data proposed in
Ref.\ \onlinecite{Post19} is not restricted to PCA, but may as well be
used in combination with nonlinear or machine learning empowered
approaches.\cite{Rohrdanz13,Glielmo21}

%
%
\subsection{Applicability and practical considerations} \label{sec:praxis}
\vspace*{-4mm}

When we apply the above introduced path separation approach to
all-atom MD simulations of ligand unbinding, various general as well
as practical problems may arise. As so often, the key issue is the
appropriate choice of coordinates to define the free energy landscape
$\Delta G(x,\vec{y})$ of the system. Most importantly, the pulling
coordinate $x$ should represent a suitable 1D reaction coordinate of the unbiased system, such that equilibrium and nonequilibrium path weights are similar. The
collective variables $\{y_j\}=\vec{y}$, on the other hand, should
allow for a low-dimensional representation of the free motion of the
system ``around'' coordinate $x$. In particular, the purpose of
coordinate $\vec{y}$ is to facilitate the detection of reaction
pathways.
As discussed above, a straightforward choice for $\vec{y}$ are the
first principal components of a PCA of protein-ligand contacts.
Representing an internal coordinate system of the protein-ligand
complex, these coordinates may
describe unbinding routes through 3D space
along which the ligand can minimize steric clashes with the
protein, thereby managing to find a low-energy exit path from the
binding pocket. Other candidates for $\vec{y}$ are protein 
backbone dihedral angles (in case that the unbinding is connected 
to a protein conformational change \cite{Amaral17,Jager22}) 
and intra-ligand hydrogen bonds (whose change affect the ligand's
conformational flexibility \cite{Bray22}).

Given these coordinates, the subsequent clustering of unbinding
trajectories into distinct reaction paths is based on the assumption
that these paths are well separated on the free energy landscape
$\Delta G(x,\vec{y})$. If the energy landscape between two paths is
flat, however, trajectories may cross between the paths and thus
hamper a clear path separation. Even for the simple 2D model
discussed in Fig.\ \ref{fig:2DHam}, we find crossing trajectories at
the beginning of both paths, which
constitute $\sim 4\,$\% of all performed simulations (see Fig.\
\SItwoDGammaVar c). In principle, we could cut crossing trajectories
into segments that are attributed to different paths. However,
similarity measures between crossing and other trajectories are
typically inconclusive, and the calculation of path energy curves
[Eq.~(\ref{eq:Pathcumulant})] using a varying number of trajectories
for each position $x$ turns out to be quite inefficient.
Considering an atomistic description of ligand unbinding, on the other
hand, crossings between various exit paths may naturally occur also
during the unbinding process. In the limiting case that the free
energy landscape is overall flat, most trajectories will cross and a
path separation ansatz does not make sense from the outset. On the
other hand, if only some smaller part of the trajectories cross, and
if we assume that they do not change the path energy curves, we may
remove these trajectories from the analysis.

Proceeding with more practical aspects of path separation, we note
that the occurrence of crossing trajectories also depends sensitively
on the pulling velocity $v$. In general, we want to choose $v$ low
enough to ensure that motions described by coordinate $\vec{y}$ have
sufficient time to relax. This corresponds to a slow adiabatic
change,\cite{Servantie03} which leaves the system virtually at
equilibrium at all positions $x$. On the other hand, if $v$ is chosen
too slow, the constraint can artificially facilitate transitions over
barriers along $\vec{y}$; a process that might not happen in the
unconstrained case. In this way, low pulling velocities may
lead to an increase of crossing trajectories. To find a balance
between these conflicting conditions, in practice we will try to find
a sweet spot of the pulling velocity, which typically is around 1\,m/s
for ligand-protein complexes (see Fig.~S4 of Ref.~\onlinecite{Wolf20}).\footnote{As practical guideline, we recommend to first perform a set of
$\sim$100 simulations for a pulling velocity of $v=1\,$m/s and check if
a path separation with contact distances is possible. If this does not
lead to a successful removal of the friction overestimation artifact
(see Fig.~1c), a similar study should be done for $v=0.1\,$m/s. If the
simulations then become dominated by crossing trajectories (see
Ref.\cite{Wolf20}, Fig. S4c), the system may be not suited for a dcTMD
analysis. We note that in all our investigations carried out so far,
\cite{Wolf18,Wolf20,Post22,Jager22,Bray22} the systems were
treatable with dcTMD, but the major challenge was finding a suitable
orthogonal coordinate $\vec{y}$.}
We note in passing that fast pulling velocities may also result in a
jamming of the ligand in an unfavorable position of the binding
pocket, which in consequence leads to a sharp peak of the constraint
force when unbinding is enforced. Since this effect would not occur at
equilibrium, however, jammed trajectories can be safely neglected.

%
%
\section{Application: The Trypsin-Benzamidine Complex} \label{Application}
\vspace*{-4mm}

The inhibitor benzamidine bound to trypsin\cite{Guillain70,
  Marquart83} is a well-established model problem to test
enhanced sampling techniques.\cite{Buch11,Plattner15,Tiwary15,
  Teo16,Votapka17,Betz19} As the unbinding process is well
characterized and occurs on a millisecond timescale, \cite{Guillain70}
the complex represents an ideal system to study the virtues and
shortcomings of the path separation scheme.

\vspace*{-4mm}
\subsection{Computational details}
\vspace*{-4mm}

MD and dcTMD simulations of the trypsin-benzamidin complex were
described in detail in Ref.\ \onlinecite{Wolf20}. In brief, all
simulations were performed using Gromacs 2018
(Ref.~\onlinecite{Abraham15}) in combination with the Amber99SB force
field \cite{Hornak06,Best09} for the protein. 
Ligand parameters were obtained from Antechamber\cite{Wang06}
with GAFF atomic parameters.\cite{Wang04} Atomic charges were obtained 
from quantum calculations at the HF/6-31G* level
using Orca\cite{Neese12} followed by RESP charge 
calculations in Multiwfn.\cite{Lu12}
The TIP3P model\cite{Jorgensen83} was used for the solvent water, and the 1.7~\AA\
crystal structure\cite{Marquart83} (PDB 3PTB) as starting condition.
dcTMD calculations were carried out using the PULL code implemented in
Gromacs. The constraint was defined via the distance between the
centers of mass of the heavy atoms of benzamidine and the C$_\alpha$
atoms of the central $\beta$-sheet of the protein. Sampling from an
equilibrium simulation at 290~K, 400 statistically independent
starting points were obtained, for which dcTMD simulations were
performed using a pulling velocity $v = 1$~m/s.
Using 400 simulations, the least populated path is taken by 52
simulations, which is typically suitable to obtain sufficiently accurate PMF and path weight estimates.\cite{Wolf18,Wolf20} The accumulated simulation time of 0.8~$\mu$s required a total wall clock time of 640 hours on a workstation with an Intel i7-3930K processor and a
NVIDIA GeForce GTX 670.

For path separation we employed a nonequilibrium PCA\cite{Post19} on
protein-ligand contacts, see Sec.\ \ref{sec:pathDetection}.
Concatenating all 400 dcTMD runs into a single pseudo-trajectory, we
determined all residues that exhibit a minimal distance
$\leq 4.5$~\AA\ between the heavy atoms of benzamidine and the
protein.\cite{Ernst15} Using these contacts and the fastpca
program,\cite{Sittel17} the nonequilibrium covariance matrix [Eq.\
\eqref{eq:CovAvNEQ}] and the associated principal components were obtained (Fig.~\SITrypPCs).
For the calculation of binding and unbinding rates, free energy and
friction fields obtained from dcTMD were employed to run unbiased
Langevin equation simulations, see the Supplementary Material for
details. To enhance the sampling, we employ the $T$-boosting technique
described in Sec.\ \ref{sec:rates}, using five independent simulations
of 200~$\mu s$ length each at 13 different temperatures between 380
and 900~K.

%
%
\subsection{Results and Discussion}
\vspace*{-4mm}

\begin{figure*}
   \centering
   \includegraphics[width=0.95\textwidth]{\dirfig/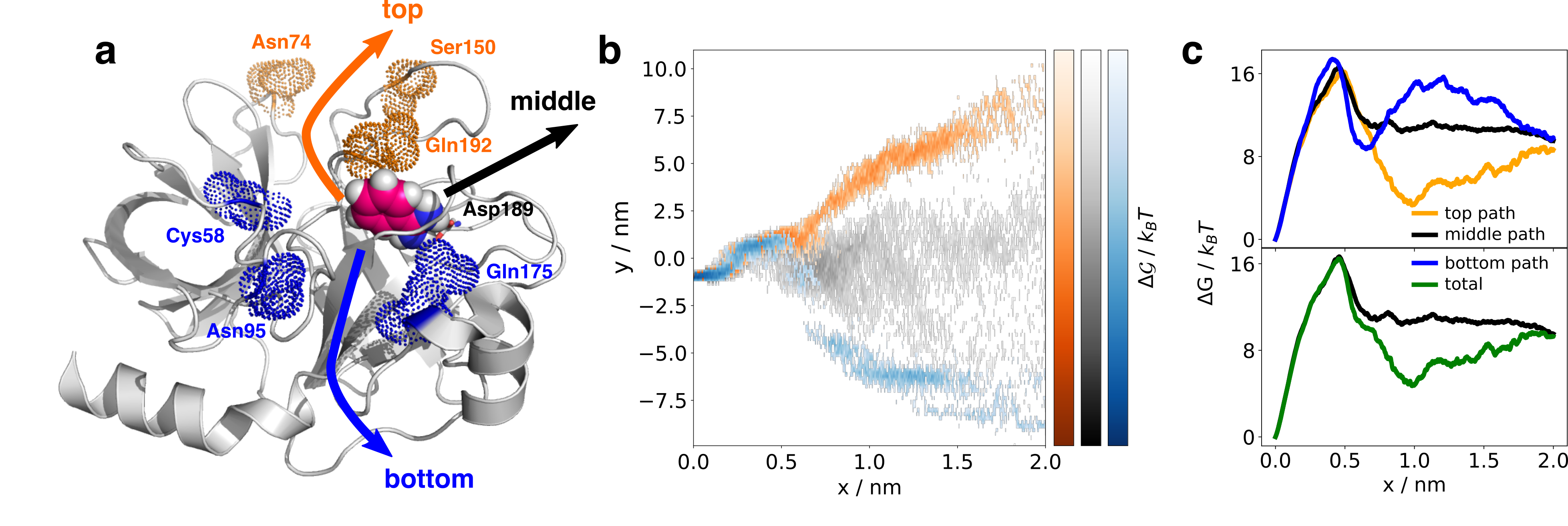}
   \caption{
     dcTMD simulations of the
     trypsin-benzamidine complex. (a) Structure of trypsin (PDB ID
     3PTB)\cite{Marquart83} showing the protein surface as cartoon
     (grey), the ligand benzamidine as van der Waals spheres
     (magenta), and residue Asp189 in sticks. Colored arrows indicate
     the direction of the three observed paths, which are
     characterized by equally colored residues displayed as
     volumes. (b) Nonequilibrium energy landscape
     $\Delta \mathcal{G}(x,y)$ indicating the three paths, represented
     as a function of the pulling coordinate $x$ and the second
     principal component $y$ of a PCA on ligand-protein contacts. (c)
     Free energy profiles along $x$ obtained by dcTMD. (Top)
     Path-specific free energy curves $\Delta G_k(x)$ with color code
     according to pathway taken. (Bottom) Comparison of the free
     energy profiles of the middle path and the combined paths
     [Eq.~(\ref{eq:pathsum4a})].}
    \label{fig:Tryp_pathsep}
\end{figure*}

Figure~\ref{fig:Tryp_pathsep} displays the structure of
trypsin and indicates the direction of the three observed dissociation
paths of benzamidine by colored arrows. Using the same colors, we also
show the resulting nonequilibrium energy landscape
$\Delta \mathcal{G}(x,y)$, represented as a function of the pulling
coordinate $x$ and the second principal component $y$ of a contact
PCA. We note that coordinate $x$ is highly correlated with the first
principal component (Fig.\ \SITrypPCs), which therefore could also be used
instead of $x$. On the other hand, coordinate $y$ encodes the
dissociation path by reporting on the direction along which
benzamidine moves over the protein surface after unbinding.
As shown in Fig.\ \SITrypPCs, the 2D representation already explains
78\% of the total variance of the unbinding process.

All unbinding simulations start with a common rupture of the
benzamidine-Asp189 salt bridge at $x\sim\,0.5$~nm, which allows the
ligand to exit the binding pocket.\cite{Wolf20} Subsequently, the
diffusion of benzamidine away from the binding site funnel is found to
follow different routes along the protein surface.  As can be derived
from the contributions of individual contacts to $y$ in
Fig.~\SITrypPCs, the ''top'' pathway (i.e., corresponding to high
values in $y$) leads to a motion along residues Asn74, Ser150 and
Gln192, which are highlighted in orange in
Fig.~\ref{fig:Tryp_pathsep}. The ''bottom'' pathway, on the other
hand, corresponds to diffusion close to residues Cys58, Asn95 and
Glu175 highlighted in blue. Finally, the ''middle'' pathway describes
unbinding directly into the bulk water. For all paths, $P(W)$ exhibits
a normal distribution, see Fig.~\SINormal.  However, we note that the
shifts of the mean $\mean{W}$ are small compared to the distributions'
widths, rendering this type of analysis difficult.

Figure~\ref{fig:Tryp_pathsep}c shows the resulting free energy
profiles $\Delta G_k(x)$ of the three paths. In line with the 2D
energy landscape, all paths show the same increase of the free energy
until the transition state at $x\sim\,0.5\,$nm, which reflects the
rupture of the benzamidine-Asp189 salt bridge. Following this main
energy barrier, the free energy of the ''top'' pathway drops to a
subsequent deep minimum at $x\sim\,1.0\,$nm, which is indicative of an
unspecific binding site on the surface of the protein involving
Tyr151. This binding site was also discussed in
Ref.~\onlinecite{Buch11}, therein named S3. The ''middle'' path, on
the other hand, rapidly levels off to the free energy value of the
unbound state. This agrees with the path leading directly into the bulk
water, which was also found in Ref.~\onlinecite{Wu16}. The ''bottom''
pathway exhibits a second minimum at $x\sim\,0.7\,$nm, 
followed by a 2\up{nd} energy barrier at $x\sim\,1.2\,$nm. This
unspecific binding site close to, e.g., Glu175 has also been observed
in Ref.~\onlinecite{Buch11} (therein named S2). Moreover, our
transition state agrees with the findings of
Ref.~\onlinecite{Tiwary15}.

\begin{table}[htbp!]
   \centering
\begin{tabular}{@{} lccccc @{}} 
      \hline
  path    & $p^\text{neq}_k$ & $p^\text{eq}_k$  & $k_{\rm off} (10^2$/s)
  & $k_{\rm on} (10^5$/sM) & $K_{\rm D} (10^{-5}$\,M)\\
  \hline
  unsep. &	--- & --- &	$20 \pm 3$ & $21 \pm 2$	& $100 \pm 20$	\\
      \hline
      top      & 0.23 	& 0.29 & $7.9 \pm 1.2$ & $0.42 \pm 0.03$  & --- \\
      middle   & 0.57  	& 0.52 & $8.0 \pm 1.0$ & $78 \pm 4$  & --- \\
      bottom \;  & 0.20   & 0.19 & $3.2 \pm 0.4$  & $45 \pm 3$ & --- \\
\hline
      total    & --- & --- & $7.1 \pm 0.9$ & $49 \pm 3$ & $15 \pm 3$ \\
      exp.\cite{Guillain70} &	---	 & --- & $6$ & $290$  &  $2$ \\
      \hline
   \end{tabular}
   \caption{
     Dissociation and association rates of the unseparated data and the
     three paths found for the trypsin-benzamidine complex (Fig.\
     \ref{fig:Tryp_pathsep}), along with their nonequilibrium and
     equilibrium weights. The resulting total rates [Eq.\
     (\ref{eq:pathsum6})] as well as the equilibrium dissociation constant
     $K_{\rm D}$ are compared to experimental data.\cite{Guillain70}
     Values in brackets denote the fitting error from
     $T$-boosting\cite{Wolf20} for path-specific rates, and results from
     error propagation for total rates.}
     \label{tab:paths}
\end{table}

Let us now turn to the analysis of the kinetics predicted for the
various pathways. Comprising the nonequilibrium and equilibrium
weights as well as the dissociation and association rates of the three
paths, Tab.\ \ref{tab:paths} reveals that the ''middle'' path is
the most populated (148 trajectories out of 400), while the two other
paths each have less than 40\,\% population. We additionally
encounter a significant number of crossing trajectories (139 out of
400), which were discarded in the analysis. Apart from being the most
accessed path, the middle pathway provides also the fastest way in and
out of the protein, and results in rates that are surprisingly similar
to the experimental findings.\cite{Guillain70} For these reasons, we
focused in our previous work\cite{Wolf20} on the main path and did not
further investigate the remaining paths. In the present more
elaborated study, however, we find that the unbinding rate
$k_{\rm off}$ is nearly identical for the three paths. This may be
related to the fact that all paths coincide on the 1D and 2D energy
landscapes in Fig.~\ref{fig:Tryp_pathsep} until the transition
state.
Concerning the binding process, we find that the top path is by two
orders of magnitude slower than the other two paths, which is due to
the deep minimum of the free energy at $x \approx 1.0$~nm. 

When we finally compare to experiment, we find that the calculated
total unbinding rates reproduces the observed rate within its error
margin, while the binding rate is underestimated by a factor of
$\sim$6. Consequently, the equilibrium dissociation constant
$K_{\rm D}=k_{\rm off}/k_{\rm on}$ is overestimated by a factor of
$\sim$6 as well. We note that this corresponds to an underestimation
of the standard free energy of binding
$\Delta G^0 = - \kb T \ln K_{\rm D}$ by $\sim$1.8~$\kb T$, which is
well below the mean error ($\sim$3~$\kb T$) of current prediction
methods of absolute binding free energy.\cite{Fu22}
Furthermore, we note that
ignoring the existence of paths and employing the free energy and
friction profiles from all 400 simulations will cause rate
estimates to deteriorate by a factor of $\sim$3 and the dissociation
constant by a factor of $\sim$7 (see Tab.~II). This comparatively small
offset comes from the fact that the friction overestimation artifact in trypsin
is particularly small around the sweet spot velocity of 1~m/s.

%
%
\section{Conclusions}

When we employ dcTMD pulling simulations to calculate free energy
profiles and reaction rates of slow processes, the practical
implementation of the method rests on the cumulant approximation of
Jarzynski's identity [Eq.\ \eqref{eq:cumulant}], which is valid given
a normal distribution of the work done on the system. By exploiting
the central limit theorem, this is found to be the case if the
integrated noise $\upint \! \d t \, \eta(t)$ associated with the
process is Gaussian distributed, see Eq.\
(\ref{eq:singleTrajWork}). While this condition was shown to be valid,
e.g., for simple and complex fluids,\cite{Post22} it typically breaks
down when the system cannot be properly described by a single 1D
pulling coordinate $x$, but requires another (in general
multidimensional) coordinate $\vec{y}$ for an appropriate
description. The problem was illustrated for a simple 2D model (Fig.\
\ref{fig:2DHam}), where the motion along $y$ gives rise to a splitting
of the 2D energy landscape into two 1D pathways. While the work
distributions of the individual paths are well approximated by a
Gaussian, their sum may exhibit a bimodal appearance, thus hampering
the direct application of the cumulant approximation.  We note that
the pathway sampling problem is not specific to dcTMD, but generally
occurs in enhanced sampling methods along a 1D biasing coordinate, see
Fig.\ \SItwoDTI\ for the case of thermodynamic integration.

As a main result of this paper, we have developed in Sec.\
\ref{sec:PathSep} the theoretical formulation underlying the
separation of the total reaction flux in several pathways. In
particular, we have established for each path $k$ the relation between
the nonequilibrium energy curves $\Delta \mathcal{G}_k(x)$ [Eq.\
\eqref{eq:pathsum4a}] and their weights $p^\text{neq}_k$ obtained from
dcTMD and their equilibrium counterparts, the free energy curves
$\Delta G_k(x)$ [Eq.\ \eqref{eq:pathsum3}] and equilibrium weights
$p^\text{eq}_k$ [Eq.\ \eqref{eq:pathsum4b}]. Combined with the
friction factors $\itGamma_k(x)$ obtained from dcTMD, the free energy
profiles $\Delta G_k(x)$ can be used to run unbiased Langevin
simulations separately for each path. This yields the reaction rates
for all paths, which are then weighted by $p^\text{eq}_k$ and added to
obtain the total rate [Eq.\ \eqref{eq:pathsum6}], thus revealing the
relative importance of the individual paths. The implementation and the
performance of the path separation procedure has been demonstrated in
detail for the 2D model in Fig.\ \ref{fig:2DHam}, where exact
reference calculations of all quantities are available.

As an application to all-atom MD data, we have considered the
(un)binding of the trypsin-benzamidine complex, which represents a
well-established model to test enhanced sampling
techniques.\cite{Buch11,Plattner15,Tiwary15, Teo16,Votapka17,Betz19}
By using a recently proposed nonequilibrium PCA on contacts between
the ligand and the protein,\cite{Post19} we have identified three
binding and unbinding pathways, which are defined in terms of their
free energy $\Delta G_k(x)$, friction $\itGamma_k(x)$ and weight
$p^\text{eq}_k$, and can be characterized via the protein residues
encountered by the ligand along the path. 
In this way, we reproduced the experimentally measured millisecond
reaction times within an error of $\sim 10\,\%$ for unbinding and
within a factor of $\sim\,$6 for binding. Moreover, we have used this
well-studied but nontrivial model problem to discuss various practical
issues of the path separation scheme, such as the assessment of the
Gaussianity of the work distribution (Fig.\ \SINormal)
and the treatment of trajectories that cross between various paths.

In future work, we want to extend the dcTMD path separation approach to
the description of more involved processes. First studies on the unbinding of
an inhibitor from the N-terminal domain of Hsp90 \cite{Wolf20}
and the ion diffusion through a membrane channel\cite{Jager22} have
indicated that the detection of independent paths with normal work
distributions represents a challenging task. This includes the
definition of suitable collective coordinates characterizing these
paths using machine learning techniques 
\cite{Ribeiro18,Ribeiro20,Bertazzo21,Badaoui22,Bray22} 
as well as the definition of distance measures between various
trajectories. \cite{Yuan17} In particular, we currently explore the
potential of MosSAIC, \cite{Diez22} a recently proposed
correlation-based feature selection procedure to classify nonequilibrium
trajectories into different paths.

%
%
\appendix
\section{Harmonic bath model}
\label{sec:appendix_harmonic_bath_model}


Zwanzig\cite{Zwanzig73} considered a system-bath Hamiltonian that can be
written as\cite{Haynes94}
\begin{align}
H &= \frac{p^2}{2m} + U(x) + H_B(x,\vec{q},\vec{p}), \\
H_B &= \sum_{i=1}^N \left[ \frac{p_{i}^2}{2m_i} + \frac{1}{2} m_i
\omega_i^2 \left(q_i-\frac{\gamma_i g(x)}{m_i \omega_i^2}  \right)^2 \right] \; .
\label{eq:hamiltonian_harmonic}
\end{align}
Here the system is described by mass $m$, coordinate $x$, momentum
$p$ and potential energy $U(x)$, the bath by masses $m_j$,
coordinates $\vec{q}=\{q_i\}$, momenta $\vec{p}=\{p_i\}$, and harmonic
frequencies $\omega_i$, and the system-bath coupling is characterized by
constants $\gamma_i$ and the function $g(x)$. 
Because the bath consists of driven harmonic oscillators with
explicitly known solutions $\{q_i(t),\,p_i(t)\}$, the equation of
motion for the system
\begin{align} \label{eq:EoMsys}
m \ddot{x} = -\frac{\partial U}{\partial x} -\frac{\partial H_B}{\partial x} 
\end{align}
can be expressed as a generalized Langevin equation of the form
\cite{Haynes94}
 \begin{align}
& m \ddot{x} = -\frac{\partial U}{\partial x} - \upint_0^t \!\! \d \tau K(t,\tau) \dot{x}(\tau) + \eta(t)\, ,\label{eq:langevin_harmonic_bath}\\
& K(t,\tau) = \sum_i \frac{ \gamma_i^2}{m_i \omega_i^2} \; g^\prime(t)  \cos(\omega_i (t-\tau)) g^\prime(\tau) \; , \label{eq:memory_harmonic_bath}\\
& \eta(t) = g^\prime(t) \sum_i \left( \gamma_i \delta q_i(0) \cos(\omega_i t) + \frac{\gamma_i \pi_{i}(0)}{m_i\omega_i}  \sin(\omega_i t) \right) \; ,
 \end{align}
where $\delta q_i(t) = q_i(t)-\frac{\gamma_i g(x(t))}{m_i
  \omega_i^2}$, $g^\prime = \frac{\partial g}{\partial x}$, and the
memory kernel $K$ and the fluctuating force $\eta$ are related by a
fluctuation-dissipation relation.
Recognizing that the free energy is given by
\begin{align}
G(x) &= -\beta^{-1} \ln \left[ \frac{1}{Z} \upint \!\! \d p \d\vec{q} \d\vec{p}
       \, \e^{ -\beta H (x, p,\vec{q}, \vec{p}) }  \right] \nonumber \\
  &= U(x) + const. \, ,
\end{align}
we find that potential energy $U(x)$ and free energy $G(x)$ coincide
up to a constant.


We now want to extend the model to the nonequilibrium regime encountered
by dcTMD, which uses the constraint $x(t)= x_0 + vt$. Considering
the associated Lagrangean $L=T-V$ of Hamiltonian $H=T+V$ in
Eq.~\eqref{eq:hamiltonian_harmonic} (with $T$ and $V$ being the
kinetic and potential energy), we can eliminate the system variable
$x$ due to the constraint. When we transform back to the Hamiltonian
description, we obtain an explicitly time-dependent Hamiltonian of the
bath degrees of freedom, 
\begin{align}
{\cal H} (t) = - \frac{1}{2} m v^2 + U(t)+ H_B(t)\, ,
\end{align}
where $U(t)$ and $H_B(t)$ denote the potential and the bath energy
evaluated at $x=x(t)$, and the first term is an irrelevant constant.
Describing driven harmonic motion, the resulting equations of
motion of the bath oscillators are again readily solved to give
\begin{align}
\delta q_i(t) &= \delta q_i^{(0)}(t) - \frac{\gamma_i v}{m_i \omega_i^2} C_i(t) \; , \\
p_{i}(t) &= p_{i}^{(0)}(t) + \frac{\gamma_i v}{\omega_i} S_i(t) \; , 
\end{align}
where $\delta q_i^{(0)}(t)= \delta q_i(0) \cos \omega_i t
+  \frac{\pi_{i}(0)}{m_i\omega_i}  \sin \omega_i t$
and $p_{i}^{(0)}(t) = -m_i \omega_i \delta q_i(0) \sin \omega_i t
+ \pi_{i}(0) \cos \omega_i t$
denote the free harmonic solutions, and
$C_i(t) = \upint_0^t \!\! \d \tau  g^\prime (\tau) \cos(\omega_i (t-\tau))$
and
$S_i(t) =\upint_0^t \!\! \d \tau g^\prime (\tau) \sin(\omega_i
(t-\tau))$.

Aiming to calculate the work performed on the system due to the
pulling along the constraint $x(t)= x_0 + vt$, we need to evaluate the
corresponding constraint force $f$. To this end, we add $f$ to the
system's equation of motion \eqref{eq:EoMsys}, and solve the equation
using $\ddot{x}=0$ resulting from the constraint. This yields
\begin{align} \label{eq:f}
  f(t) = \left. \frac{\partial U}{\partial x}\right|_{x(t)}
  \! - g^\prime(t) \sum_i \gamma_i \, \delta q_i(t) \; ,
\end{align}
from which the work $W$ [Eq.~\eqref{eq:Work}] evaluates as
\begin{align}
W(t) &= \Delta U (x(t)) + v^2 \sum_i \frac{\gamma_i^2}{m_i
       \omega_i^2} I_i(t) \nonumber \\
     & \quad - \; v \sum_i \left[ \gamma_i \delta q_i(0) \widehat{C}_i(t)
       + \frac{\gamma_i \pi_{i}(0)}{m_i \omega_i}  \widehat{S}_i(t) \right],
\end{align}
where $I_i(t) = \frac{1}{2} \left( \widehat{C}_i^2(t) + \widehat{S}_i^2(t) \right)$,
$\widehat{C}_i(t) = S_i(t) \sin\omega_i t  + C_i(t) \cos \omega_i t$, 
and $\widehat{S}_i(t) = C_i(t) \sin \omega_i t  - S_i(t) \cos \omega_i t $.
To obtain the distribution of the work, we construct the characteristic function
\begin{align}
\psi_{W(x)}(\alpha) &= \EV{ \e^{\im \alpha W(x)} }  \\
                    &= \exp\left\{ \im \alpha \left[\Delta U(x)
+ v^2 \sum_i \frac{\gamma_i^2}{m_i \omega_i^2} I_i(t) \right] \right. \nonumber \\
& \quad \quad \quad \left. -\frac{\alpha^2}{2} \left[ \frac{2 v^2}{\beta} 
\sum_i \frac{\gamma_i^2}{m_i \omega_i^2} I_i(t)  \right] \right\} \; , \label{eq:charfunct}
\end{align}
where $\EV{\ldots}$ denotes the Boltzmann average over the initial
bath degrees of freedom. Equation \eqref{eq:charfunct} represents the
characteristic function of a Gaussian with two nonzero first
cumulants. From the second cumulant, we obtain the mean dissipated
energy
\begin{align}
\EV{W_\text{diss}(t)} &= v^2 \sum_i \frac{\gamma_i^2}{m_i \omega_i^2}I_i(t)
 \label{eq:harmonic_bath_diss} \\
              &=  \sum_i \frac{ \EV{ p_{i}^2(t) } }{2m_i} +
                \frac{1}{2} m_i \omega_i^2 \EV{ \delta q_i^2 (t) } 
                 - N/\beta \, , \nonumber
\end{align}
which reflects the increase of the bath energy.

Finally, we note that the nonequilibrium generalized Langevin equation
\eqref{eq:GLE} describing dcTMD is directly obtained from the system's
equation of motion \eqref{eq:EoMsys} by adding the constraint force
$f$. Using Eq.\ \eqref{eq:f}, the constraint force autocorrelation
is evaluated as
\begin{align}
\beta \EV{ \delta f(t_1) \delta f(t_2) } =  \sum_i
  \frac{\gamma_i^2}{m_i \omega_i^2} g^\prime(t_1)
  \cos \omega_i(t_1\!-\!t_2)\, g^\prime(t_2) \, ,
\label{eq:harmonic_ffacf}
\end{align}
which equals the expression for the memory kernel $K$ in
Eq.~\eqref{eq:memory_harmonic_bath}, thus describing a
dissipation-fluctuation relation. Hence, for the harmonic model, the
memory kernel measured via a constrained simulation is of the same
form as the kernel obtained from the free motion.

%
%
\subsection*{Author's contributions}
\vspace*{-4mm}

All authors contributed equally to this work.

\subsection*{Acknowledgments}
\vspace*{-4mm}

The authors thank Simon Bray, Miriam J\"ager, Moritz Sch\"affler and
Viktor T\"anzel for helpful comments and discussions.  This work has
been supported by the Deutsche Forschungsgemeinschaft (DFG) within the
framework of the Research Unit FOR 5099 ''Reducing complexity of
nonequilibrium'' (project No.~431945604), the High Performance and
Cloud Computing Group at the Zentrum f\"ur Datenverarbeitung of the
University of T\"ubingen, the state of Baden-W\"urttemberg through
bwHPC and the DFG through grant no INST 37/935-1 FUGG (RV bw16I016),
the Black Forest Grid Initiative, and the Freiburg Institute for
Advanced Studies (FRIAS) of the Albert-Ludwigs-University Freiburg.

\subsection*{Data availability}
\vspace*{-4mm}

Python scripts for dcTMD data evaluation as well as the Jupyter
notebook used for the 2D model calculations can be obtained at
\url{https://github.com/moldyn}. Here also a tutorial for the usage of
dcTMD with the trypsin-benzamidine complexes including simulation
start structure, topologies and Gromacs command input files is
available, as well as the \url{fastpca} program package for
nonequilibrium PCA.

%
%


\begin{thebibliography}{10}

\bibitem{Chipot07}
C.~Chipot and A.~Pohorille,
\newblock {\em Free Energy Calculations},
\newblock Springer, Berlin, 2007.

\bibitem{Wales03}
D.~J. Wales,
\newblock {\em Energy Landscapes},
\newblock Cambridge University Press, Cambridge, 2003.

\bibitem{Faccioli06}
P.~Faccioli, M.~Sega, F.~Pederiva, and H.~Orland,
\newblock Dominant pathways in protein folding,
\newblock Phys. Rev. Lett. {\bf 97}, 108101 (2006).

\bibitem{Dellago16}
C.~Dellago and P.~Bolhuis,
\newblock Transition path sampling and other advanced simulation techniques for
  rare events,
\newblock Adv. Polymer Sci. {\bf 221} (2016).

\bibitem{Zuckerman17}
D.~M. Zuckerman and L.~T. Chong,
\newblock Weighted ensemble simulation: Review of methodology, applications,
  and software,
\newblock Annu. Rev. Biophys. {\bf 46}, 43  (2017).

\bibitem{Capelli19}
R.~Capelli, P.~Carloni, and M.~Parrinello,
\newblock {Exhaustive Search of Ligand Binding Pathways via Volume-Based
  Metadynamics},
\newblock J. Phys. Chem. Lett. , 3495 (2019).

\bibitem{Capelli19a}
R.~Capelli, A.~Bochicchio, G.~Piccini, R.~Casasnovas, P.~Carloni, and
  M.~Parrinello,
\newblock {Chasing the Full Free Energy Landscape of Neuroreceptor/Ligand
  Unbinding by Metadynamics Simulations.},
\newblock J. Chem. Theory Comput. {\bf 15}, 3354 (2019).

\bibitem{Rydzewski19}
J.~Rydzewski and O.~Valsson,
\newblock {Finding multiple reaction pathways of ligand unbinding},
\newblock J. Chem. Phys. {\bf 150}, 221101 (2019).

\bibitem{Bianciotto21}
M.~Bianciotto, P.~Gkeka, D.~B. Kokh, R.~C. Wade, and H.~Minoux,
\newblock {Contact Map Fingerprints of Protein-Ligand Unbinding Trajectories
  Reveal Mechanisms Determining Residence Times Computed from Scaled Molecular
  Dynamics.},
\newblock J. Chem. Theory Comput. {\bf 17}, 6522 (2021).

\bibitem{NunesAlves21}
A.~Nunes-Alves, D.~B. Kokh, and R.~C. Wade,
\newblock {Ligand unbinding mechanisms and kinetics for T4 lysozyme mutants
  from $\tau$RAMD simulations},
\newblock Curr. Res. Struct. Biol. {\bf 3}, 106 (2021).

\bibitem{Henin22}
J.~Hénin, T.~Lelièvre, M.~R. Shirts, O.~Valsson, and L.~Delemotte,
\newblock Enhanced sampling methods for molecular dynamics simulations,
\newblock LiveCoMS {\bf 4}, 1583 (2022).

\bibitem{Wolf18}
S.~Wolf and G.~Stock,
\newblock Targeted molecular dynamics calculations of free energy profiles
  using a nonequilibrium friction correction,
\newblock J. Chem. Theory Comput. {\bf 14}, 6175  (2018).

\bibitem{Wolf20}
S.~Wolf, B.~Lickert, S.~Bray, and G.~Stock,
\newblock Multisecond ligand dissociation dynamics from atomistic simulations,
\newblock Nat. Commun. {\bf 11}, 2918 (2020).

\bibitem{Post22}
M.~Post, S.~Wolf, and G.~Stock,
\newblock Molecular origin of driving-dependent friction in fluids,
\newblock J. Chem. Theory Comput. {\bf 18}, 2816 – 2825 (2022).

\bibitem{Tiwary15}
P.~Tiwary, V.~Limongelli, M.~Salvalaglio, and M.~Parrinello,
\newblock {Kinetics of protein{\textendash}ligand unbinding: Predicting
  pathways, rates, and rate-limiting steps},
\newblock Proc. Natl. Acad. Sci. USA {\bf 112}, E386 (2015).

\bibitem{Plattner15}
N.~Plattner and F.~No{\'e},
\newblock {Protein conformational plasticity and complex ligand-binding
  kinetics explored by atomistic simulations and Markov models},
\newblock Nat. Commun. {\bf 6}, 7653 (2015).

\bibitem{Teo16}
I.~Teo, C.~G. Mayne, K.~Schulten, and T.~Lelievre,
\newblock {Adaptive Multilevel Splitting Method for Molecular Dynamics
  Calculation of Benzamidine-Trypsin Dissociation Time},
\newblock J. Chem. Theory Comput. {\bf 12}, 2983 (2016).

\bibitem{Jarzynski97}
C.~Jarzynski,
\newblock Nonequilibrium equality for free energy differences,
\newblock Phys. Rev. Lett. {\bf 78}, 2690 (1997).

\bibitem{Hendrix01}
D.~A. Hendrix and C.~Jarzynski,
\newblock {A ``fast growth'' method of computing free energy differences},
\newblock J. Chem. Phys. {\bf 114}, 5974 (2001).

\bibitem{Dellago14}
C.~Dellago and G.~Hummer,
\newblock Computing equilibrium free energies using non-equilibrium molecular
  dynamics,
\newblock Entropy {\bf 16}, 41 (2014).

\bibitem{Ytreberg04}
F.~M. Ytreberg and D.~M. Zuckerman,
\newblock {Efficient use of nonequilibrium measurement to estimate free energy
  differences for molecular systems},
\newblock J. Comput. Chem. {\bf 25}, 1749 (2004).

\bibitem{Echeverria12}
I.~Echeverria and L.~M. Amzel,
\newblock {Estimation of Free-Energy Differences from Computed Work
  Distributions: An Application of Jarzynski{\textquoteright}s Equality},
\newblock J. Phys. Chem. B {\bf 116}, 10986 (2012).

\bibitem{Billingsley95}
P.~Billingsley,
\newblock {\em Probability and Measure},
\newblock Wiley, 1995.

\bibitem{Jager22}
M.~J{\"a}ger, T.~Koslowski, and S.~Wolf,
\newblock {Predicting Ion Channel Conductance via Dissipation-Corrected
  Targeted Molecular Dynamics and Langevin Equation Simulations},
\newblock J. Chem. Theory Comput. {\bf 18}, 494 (2022).

\bibitem{Doudou09}
S.~Doudou, N.~A. Burton, and R.~H. Henchman,
\newblock {Standard Free Energy of Binding from a One-Dimensional Potential of
  Mean Force},
\newblock J. Chem. Theory Comput. {\bf 5}, 909 (2009).

\bibitem{Post19}
M.~Post, S.~Wolf, and G.~Stock,
\newblock Principal component analysis of nonequilibrium molecular dynamics
  simulations,
\newblock J. Chem. Phys. {\bf 150}, 204110 (2019).

\bibitem{Guillain70}
F.~Guillain and D.~Thusius,
\newblock {Use of proflavine as an indicator in temperature-jump studies of the
  binding of a competitive inhibitor to trypsin},
\newblock J. Am. Chem. Soc. {\bf 92}, 5534 (1970).

\bibitem{Marquart83}
M.~Marquart, J.~Walter, J.~Deisenhofer, W.~Bode, and R.~Huber,
\newblock {The geometry of the reactive site and of the peptide groups in
  trypsin, trypsinogen and its complexes with inhibitors},
\newblock Acta Crystallogr. B {\bf 39}, 480 (1983).

\bibitem{Buch11}
I.~Buch, T.~Giorgino, and G.~De~Fabritiis,
\newblock Complete reconstruction of an enzyme-inhibitor binding process by
  molecular dynamics simulations,
\newblock Proc. Natl. Acad. Sci. USA {\bf 108}, 10184 (2011).

\bibitem{Votapka17}
L.~W. Votapka, B.~R. Jagger, A.~Heyneman, and R.~E. Amaro,
\newblock {SEEKR: simulation enabled estimation of kinetic rates, a
  computational tool to estimate molecular kinetics and its application to
  trypsin{\textendash}benzamidine binding},
\newblock J. Phys. Chem. B {\bf 121}, 3597 (2017).

\bibitem{Betz19}
R.~M. Betz and R.~O. Dror,
\newblock {How Effectively Can Adaptive Sampling Methods Capture Spontaneous
  Ligand Binding?},
\newblock J. Chem. Theory Comput. {\bf 15}, 2053 (2019).

\bibitem{Schlitter94}
J.~Schlitter, M.~Engels, and P.~{Kr\"uger},
\newblock Targeted molecular dynamics - a new approach for searching pathways
  of conformational transitions,
\newblock J. Mol. Graph. {\bf 12}, 84 (1994).

\bibitem{Harris20}
C.~R. Harris et~al.,
\newblock Array programming with {NumPy},
\newblock Nature {\bf 585}, 357 (2020).

\bibitem{Virtanen20}
P.~Virtanen et~al.,
\newblock {SciPy 1.0: fundamental algorithms for scientific computing in
  Python},
\newblock Nat. Methods {\bf 17}, 261 (2020).

\bibitem{McKinney10}
W.~McKinney,
\newblock Data structures for statistical computing in python.,
\newblock in {\em Proceedings of the 9th Python in Science Conference.}, edited
  by S.~van~der Walt and J.~Millman, pages 51--56, 2010.

\bibitem{hunter07}
J.~D. Hunter,
\newblock {Matplotlib: A 2D graphics environment},
\newblock Comput. Sci. Eng. {\bf 9}, 90 (2007).

\bibitem{gnuplot}
J.~Racine,
\newblock {gnuplot 4.0: a portable interactive plotting utility},
\newblock J. Appl. Econ. {\bf 21}, 133 (2006).

\bibitem{Kluyver16}
T.~Kluyver et~al.,
\newblock {Jupyter Notebooks-a publishing format for reproducible computational
  workflows.},
\newblock in {\em Positioning and Power in Academic Publishing}, edited by
  F.~Loizides and B.~Schmidt, pages 87--90, IOP Press, 2016.

\bibitem{Zwanzig73}
R.~Zwanzig,
\newblock Nonlinear generalized {Langevin} equations,
\newblock J. Stat. Phys. {\bf 9}, 215  (1973).

\bibitem{Sorensen00}
M.~R. S{\o}rensen and A.~F. Voter,
\newblock Temperature-accelerated dynamics for simulation of infrequent events,
\newblock J. Chem. Phys. {\bf 112}, 9599 (2000).

\bibitem{Henderson06}
A.~R. Henderson,
\newblock {Testing experimental data for univariate normality.},
\newblock Clin. Chim. Acta {\bf 366}, 112 (2006).

\bibitem{Logan96}
S.~R. Logan and D.~Wilmer,
\newblock {\em Fundamentals of chemical kinetics}, volume~10,
\newblock Longman London, 1996.

\bibitem{Noe17}
F.~No{\'e} and C.~Clementi,
\newblock Collective variables for the study of long-time kinetics from
  molecular trajectories: theory and methods,
\newblock Curr. Opin. Struct. Biol. {\bf 43}, 141  (2017).

\bibitem{Sittel18}
F.~Sittel and G.~Stock,
\newblock Perspective: Identification of collective coordinates and metastable
  states of protein dynamics,
\newblock J. Chem. Phys. {\bf 149}, 150901 (2018).

\bibitem{Ernst15}
M.~Ernst, F.~Sittel, and G.~Stock,
\newblock Contact- and distance-based principal component analysis of protein
  dynamics,
\newblock J. Chem. Phys. {\bf 143}, 244114 (2015).

\bibitem{Amadei93}
A.~Amadei, A.~B.~M. Linssen, and H.~J.~C. Berendsen,
\newblock Essential dynamics of proteins,
\newblock Proteins {\bf 17}, 412 (1993).

\bibitem{Altis08}
A.~Altis, M.~Otten, P.~H. Nguyen, R.~Hegger, and G.~Stock,
\newblock Construction of the free energy landscape of biomolecules via
  dihedral angle principal component analysis,
\newblock J. Chem. Phys. {\bf 128}, 245102 (2008).

\bibitem{Hummer01}
G.~Hummer, A.~E. Garcia, and S.~Garde,
\newblock Helix nucleation kinetics from molecular simulations in explicit
  water,
\newblock Proteins {\bf 42}, 77 (2001).

\bibitem{Rohrdanz13}
M.~A. Rohrdanz, W.~Zheng, and C.~Clementi,
\newblock Discovering mountain passes via torchlight: Methods for the
  definition of reaction coordinates and pathways in complex macromolecular
  reactions,
\newblock Annu. Rev. Phys. Chem. {\bf 64}, 295 (2013).

\bibitem{Glielmo21}
A.~Glielmo, B.~E. Husic, A.~Rodriguez, C.~Clementi, F.~No{\'e}, and A.~Laio,
\newblock Unsupervised learning methods for molecular simulation data,
\newblock Chem. Rev. {\bf 121}, 9722 (2021).

\bibitem{Amaral17}
M.~Amaral, D.~B. Kokh, J.~Bomke, A.~Wegener, H.~P. Buchstaller, H.~M.
  Eggenweiler, P.~Matias, C.~Sirrenberg, R.~C. Wade, and M.~Frech,
\newblock {Protein conformational flexibility modulates kinetics and
  thermodynamics of drug binding.},
\newblock Nat. Commun. {\bf 8}, 2276 (2017).

\bibitem{Bray22}
S.~Bray, V.~T{\"a}nzel, and S.~Wolf,
\newblock {Ligand Unbinding Pathway and Mechanism Analysis Assisted by Machine
  Learning and Graph Methods},
\newblock J. Chem. Inf. Model. {\bf 62}, 4591 (2022).

\bibitem{Servantie03}
J.~Servantie and P.~Gaspard,
\newblock {Methods of calculation of a friction coefficient: application to
  nanotubes},
\newblock Phys. Rev. Lett. {\bf 91}, 185503 (2003).

\bibitem{Note1}
As practical guideline, we recommend to first perform a set of $\sim $100
  simulations for a pulling velocity of $v=1\protect \tmspace +\thinmuskip
  {.1667em}$m/s and check if a path separation with contact distances is
  possible. If this does not lead to a successful removal of the friction
  overestimation artifact (see Fig.~1c), a similar study should be done for
  $v=0.1\protect \tmspace +\thinmuskip {.1667em}$m/s. If the simulations then
  become dominated by crossing trajectories (see Ref.\cite {Wolf20}, Fig. S4c),
  the system may be not suited for a dcTMD analysis. We note that in all our
  investigations carried out so far, \cite
  {Wolf18,Wolf20,Post22,Jager22,Bray22} the systems were treatable with dcTMD,
  but the major challenge was finding a suitable orthogonal coordinate
  ${\protect \boldsymbol {y}}$.

\bibitem{Abraham15}
M.~J. Abraham, T.~Murtola, R.~Schulz, S.~Pall, J.~C. Smith, B.~Hess, and
  E.~Lindahl,
\newblock Gromacs: High performance molecular simulations through multi-level
  parallelism from laptops to supercomputers,
\newblock SoftwareX {\bf 1}, 19  (2015).

\bibitem{Hornak06}
V.~Hornak, R.~Abel, A.~Okur, B.~Strockbine, A.~Roitberg, and C.~Simmerling,
\newblock Comparison of multiple {Amber} force fields and development of
  improved protein backbone parameters,
\newblock Proteins {\bf 65}, 712 (2006).

\bibitem{Best09}
R.~B. Best and G.~Hummer,
\newblock Optimized molecular dynamics force fields applied to the helix-coil
  transition of polypeptides,
\newblock J. Phys. Chem. B {\bf 113}, 9004 (2009).

\bibitem{Wang06}
J.~Wang and R.~{Br\"uschweiler},
\newblock {2D} entropy of discrete molecular ensembles,
\newblock J. Chem. Theory Comput. {\bf 2}, 18 (2006).

\bibitem{Wang04}
J.~M. Wang, R.~M. Wolf, J.~W. Caldwell, P.~A. Kollman, and D.~A. Case,
\newblock {Development and testing of a general amber force field},
\newblock J. Comput. Chem. {\bf 25}, 1157 (2004).

\bibitem{Neese12}
F.~Neese,
\newblock {The ORCA program system},
\newblock WIREs Comput. Mol. Sci. {\bf 2}, 73 (2012).

\bibitem{Lu12}
T.~Lu and F.~Chen,
\newblock {Multiwfn: A multifunctional wavefunction analyzer},
\newblock J. Comput. Chem. {\bf 33}, 580 (2012).

\bibitem{Jorgensen83}
W.~L. Jorgensen, J.~Chandrasekhar, J.~D. Madura, R.~W. Impey, and M.~Klein,
\newblock Comparison of simple potential functions for simulating liquid water,
\newblock J. Chem. Phys. {\bf 79}, 926 (1983).

\bibitem{Sittel17}
F.~Sittel, T.~Filk, and G.~Stock,
\newblock Principal component analysis on a torus: Theory and application to
  protein dynamics,
\newblock J. Chem. Phys. {\bf 147}, 244101 (2017).

\bibitem{Wu16}
H.~Wu, F.~Paul, C.~Wehmeyer, and F.~No{\'e},
\newblock {Multiensemble Markov models of molecular thermodynamics and
  kinetics},
\newblock Proc. Natl. Acad. Sci. USA {\bf 113}, E3221 (2016).

\bibitem{Fu22}
H.~Fu, Y.~Zhou, X.~Jing, X.~Shao, and W.~Cai,
\newblock {Meta-Analysis Reveals That Absolute Binding Free-Energy Calculations
  Approach Chemical Accuracy},
\newblock J. Med. Chem. {\bf 65}, 12970 (2022).

\bibitem{Ribeiro18}
J.~M.~L. Ribeiro, P.~Bravo, Y.~Wang, and P.~Tiwary,
\newblock Reweighted autoencoded variational bayes for enhanced sampling
  (rave),
\newblock J. Chem. Phys. {\bf 149}, 072301 (2018).

\bibitem{Ribeiro20}
J.~M.~L. Ribeiro, D.~Provasi, and M.~Filizola,
\newblock {A combination of machine learning and infrequent metadynamics to
  efficiently predict kinetic rates, transition states, and molecular
  determinants of drug dissociation from G protein-coupled receptors},
\newblock J. Chem. Phys. {\bf 153}, 124105 (2020).

\bibitem{Bertazzo21}
M.~Bertazzo, D.~Gobbo, S.~Decherchi, and A.~Cavalli,
\newblock {Machine Learning and Enhanced Sampling Simulations for Computing the
  Potential of Mean Force and Standard Binding Free Energy.},
\newblock J. Chem. Theory Comput. {\bf 17}, 5287 (2021).

\bibitem{Badaoui22}
M.~Badaoui et~al.,
\newblock {Combined Free-Energy Calculation and Machine Learning Methods for
  Understanding Ligand Unbinding Kinetics.},
\newblock J. Chem. Theory Comput. {\bf 18}, 2543 (2022).

\bibitem{Yuan17}
G.~Yuan, P.~Sun, J.~Zhao, and D.~Li,
\newblock A review of moving object trajectory clustering algorithms,
\newblock Artif. Intel.l Rev. {\bf 47}, 123 – 144 (2017).

\bibitem{Diez22}
G.~Diez, D.~Nagel, and G.~Stock,
\newblock Correlation-based feature selection to identify functional dynamics
  in proteins,
\newblock J. Chem. Theory Comput. {\bf 18}, 5079 – 5088 (2022).

\bibitem{Haynes94}
G.~R. Haynes, G.~A. Voth, and E.~Pollak,
\newblock A theory for the activated barrier crossing rate constant in systems
  influenced by space and time dependent friction,
\newblock J. Chem. Phys. {\bf 101}, 7811 (1994).

\end{thebibliography}

\begin{thebibliography}{1}

\bibitem{Post19}
M.~Post, S.~Wolf, and G.~Stock,
\newblock Principal component analysis of nonequilibrium molecular dynamics
  simulations,
\newblock J. Chem. Phys. {\bf 150}, 204110 (2019).

\end{thebibliography}


\clearpage
\onecolumngrid

\renewcommand{\thepage}{\arabic{chapter}.\arabic{page}}  
\renewcommand{\thesection}{\arabic{chapter}.\arabic{section}}   
\renewcommand{\thetable}{\arabic{chapter}.\arabic{table}}   
\renewcommand{\thefigure}{\arabic{chapter}.\arabic{figure}}
\renewcommand{\thepage}{S\arabic{page}}  
\renewcommand{\thesection}{S\arabic{section}}   
\renewcommand{\thetable}{S\arabic{table}}   
\renewcommand{\thefigure}{S\arabic{figure}}

\setcounter{page}{1}    
\setcounter{section}{0}    
\setcounter{figure}{0}    
\setcounter{enumiv}{0} 

\section*{Supplementary Material}

\section{2D ligand-protein dissociation model}
	
\begin{figure}[htb!]
		\centering
		\includegraphics[width=0.55\textwidth]{\dirfig/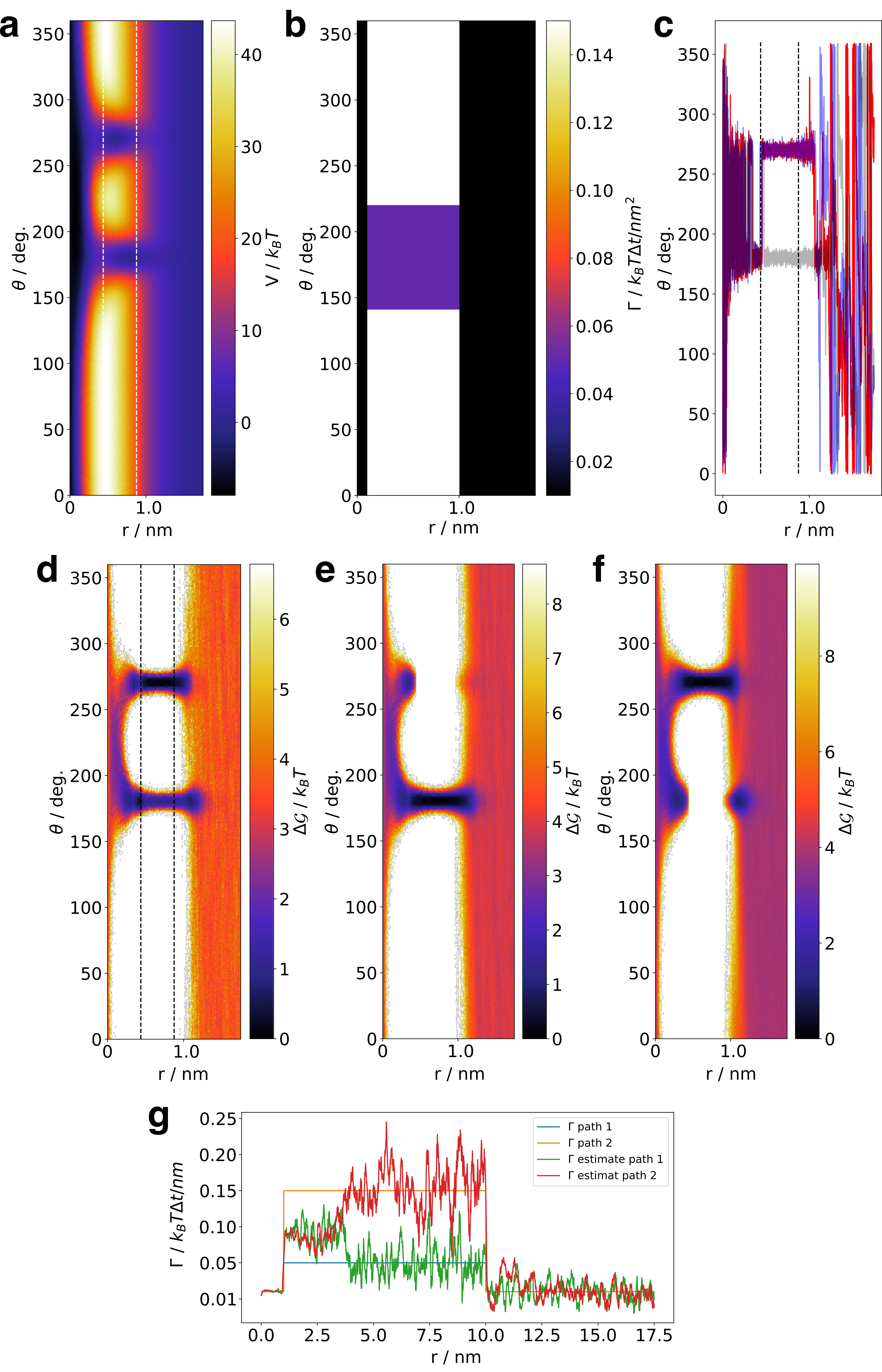}
                \caption{\baselineskip4mm 2D model of ligand-protein
                  dissociation. (a) Potential energy landscape
                  $V(r,\theta)$ including borders for path
                  discrimination at $r \in [0.44, 0.875]$~nm. (b)
                  Friction field $\itGamma (r,\theta)$. (c)
                  Representative examples of crossing trajectories,
                  exhibiting jumps at the beginning of the path
                  separation domain. (d) Sampling of the density
                  obtained from 5000 trajectories for the full system,
                  and for (e) path\,1 and (f) path\,2. (g) Friction
                  estimate $\itGamma$ from targeted MD simulations,
                  smoothed by a window average filter with 0.0125~nm
                  width.  Before path separation, $\itGamma$ is
                  estimated as a mean of the two true path friction
                  coefficients. After path separation, estimates of
                  $\itGamma$ fluctuate around the correct value.}
		\label{fig:2DHam_detail}
	\end{figure}

	We constructed a 2D potential energy function
	\begin{align}
		\beta V(r_1,r_2) &= 140.3 \times N(r_1,0,0.5) N(r_2,0,0.5) \nonumber \\
		&\quad - 31.3 \times N(r_1,0,0.25) N(r_2,0,0.25) \nonumber \\
		& \quad - 11.0 \times N(r_1,0.5,0.3) N(r_2,0,0.15) \nonumber \\
		& \quad - 9.8 \times N(r_1,0,0.15) N(r_2,0.5,0.25) \, , \quad \text{with}  \label{eq:2DHam} \\
		N(r,\mu,\sigma) &= \frac{1}{\sqrt{2\pi \sigma^2}} \exp\left\{- \frac{(r-\mu)^2}{2\sigma^2}\right\} \; , \nonumber
	\end{align}
        for $r_1^2+r_2^2 < 1.75^2$ (lengths are in units of nm), and
        $V(r_1,r_2) = +\infty$ elsewise, see Fig.~1a of the main text.
        This potential represents a 2D ''flatland'' globular protein
        in implicit solvent with a central binding cavity and two
        unbinding pathways 1 ($\theta \sim 180^\circ$) and 2
        ($\theta \sim 270^\circ$), see Fig.\ \ref{fig:2DHam_detail}a.
	
	
\subsection{MC/LE simulations}
	
For constraint simulations with anisotropic friction profiles
$\itGamma$, we used a combined Monte Carlo/Langevin equation approach
instead of a direct integration of the Langevin equation, as it is
more robust, does not require to define a suitable system mass or
range of $\itGamma$ to perform a stable numerical integration of a
Langevin equation, and allows to tune $\itGamma$ for demonstration
purposes in such a way that $P(W)$ becomes clearly
multimodal. Introducing polar $(r,\theta)$ coordinates, the potential
$V(r,\theta)$ as shown in Fig.~\SItwoDGammaVar a. We also display the
position-dependent friction profile (Fig.~\SItwoDGammaVar b), with
$\itGamma=$~0.05 $\kb T\, \Delta t$~nm$^{-2}$ for $r \in [0.1,1.0]$~nm
along path~1, $\itGamma=$~0.15 $\kb T\, \Delta t$~nm$^{-2}$ for path~2
and elsewhere between 0.1~nm and 1.0~nm, and $\itGamma=$~0.01
$\kb T\, \Delta t$~nm$^{-2}$ for $r \notin [0.1,1.0]$~nm, i.e., at all
other positions. The motion along $r= r_0 + vt$ is defined via the
constant velocity $v$. For motions in $\theta$, we used a step size of
$\Delta \theta = 100^{\circ}\, \mathcal{R} / r$ with a random number
$\mathcal{R}$ drawn from a normal distribution with $\mu = 0$ and
$\sigma = 1$, and a Jacobian term of $1/r$ to compare MC calculations
in $(r_1,r_2)$ and in $(r,\theta)$.  From the position along
$(r, \theta)$, force profiles $f$ were constructed according to
\begin{align} 
		f(r,\theta) = -\left( \frac{\partial V}{\partial r} \right)_\theta - v \itGamma(r,\theta) + \sqrt{2 \itGamma(r,\theta) \kb T} \, \mathcal{R}.
		\label{eq:LE} 
\end{align}

Sampling along $\theta$ was carried out via MC simulations using the
Metropolis acceptance criterion.  5000 constraint simulations were
initialized at randomized positions along $\theta$ and $r =
0.01$~nm. Pulling was carried out with a velocity
$v = 2.5 \cdot 10^{-4}$~nm / MC step for constraint pulling, and
$v = 0$ for thermodynamic integration (TI) calculations.  Trajectory
work profiles $W$ from constraint pulling were calculated from
$W = \int \d r \, f(r)$. For TI calculations, mean forces
$\mean{f(r_i)}_\theta$ at position $r_i$ were calculated as
$\mean{f(r_i)}_\theta = \int \d \theta \, f( r_i, \theta )\, P(\theta | r_i )$,
and free energies estimated as
$\Delta G_{\rm TI} \approx \sum_i \mean{f(r_i)}_\theta \, \Delta r$.

	
	
\clearpage
	
	
	\begin{figure}[htb!]
		\centering
		\includegraphics[width=0.3\textwidth]{\dirfig/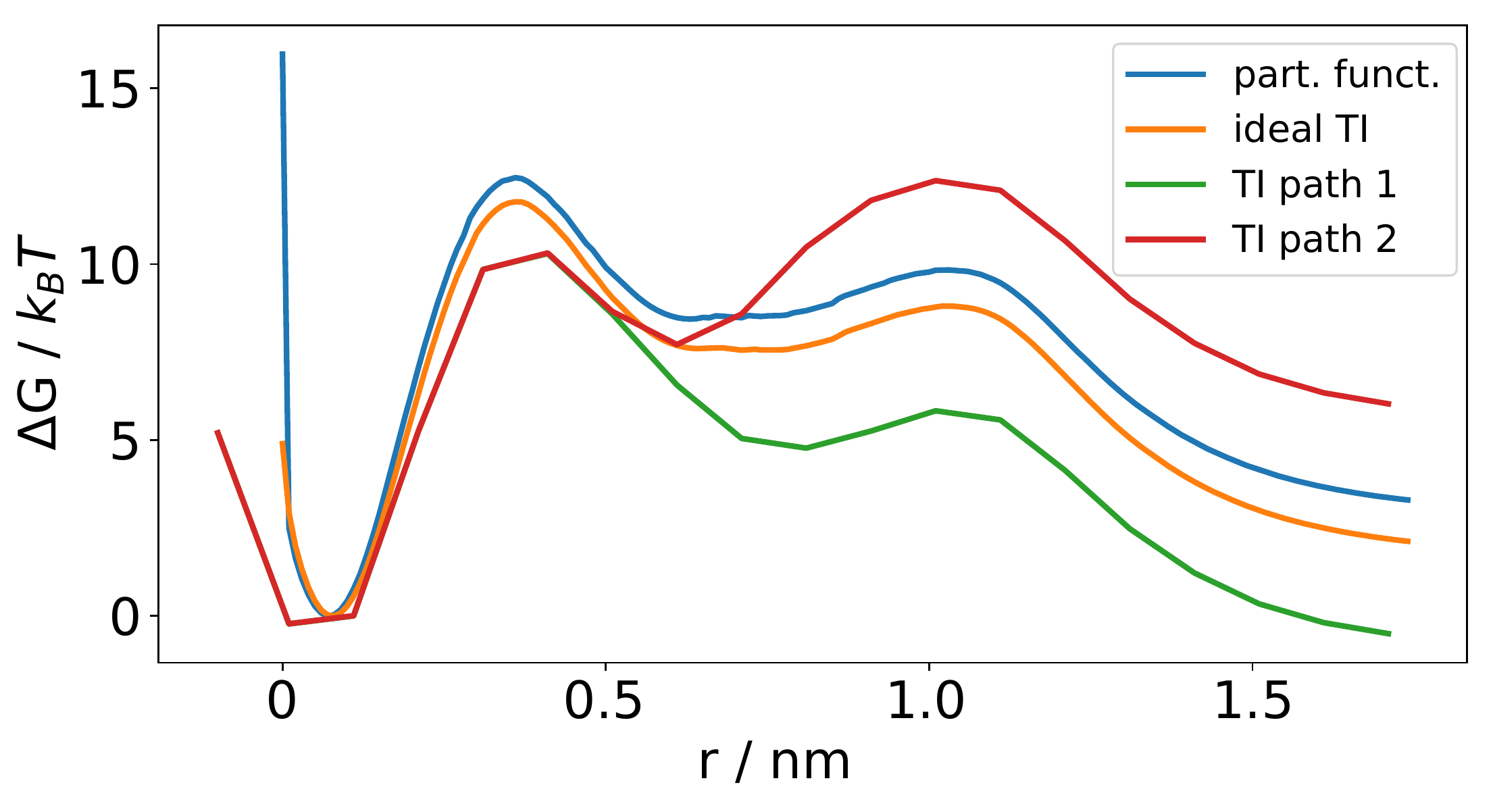}
		\includegraphics[width=0.3\textwidth]{\dirfig/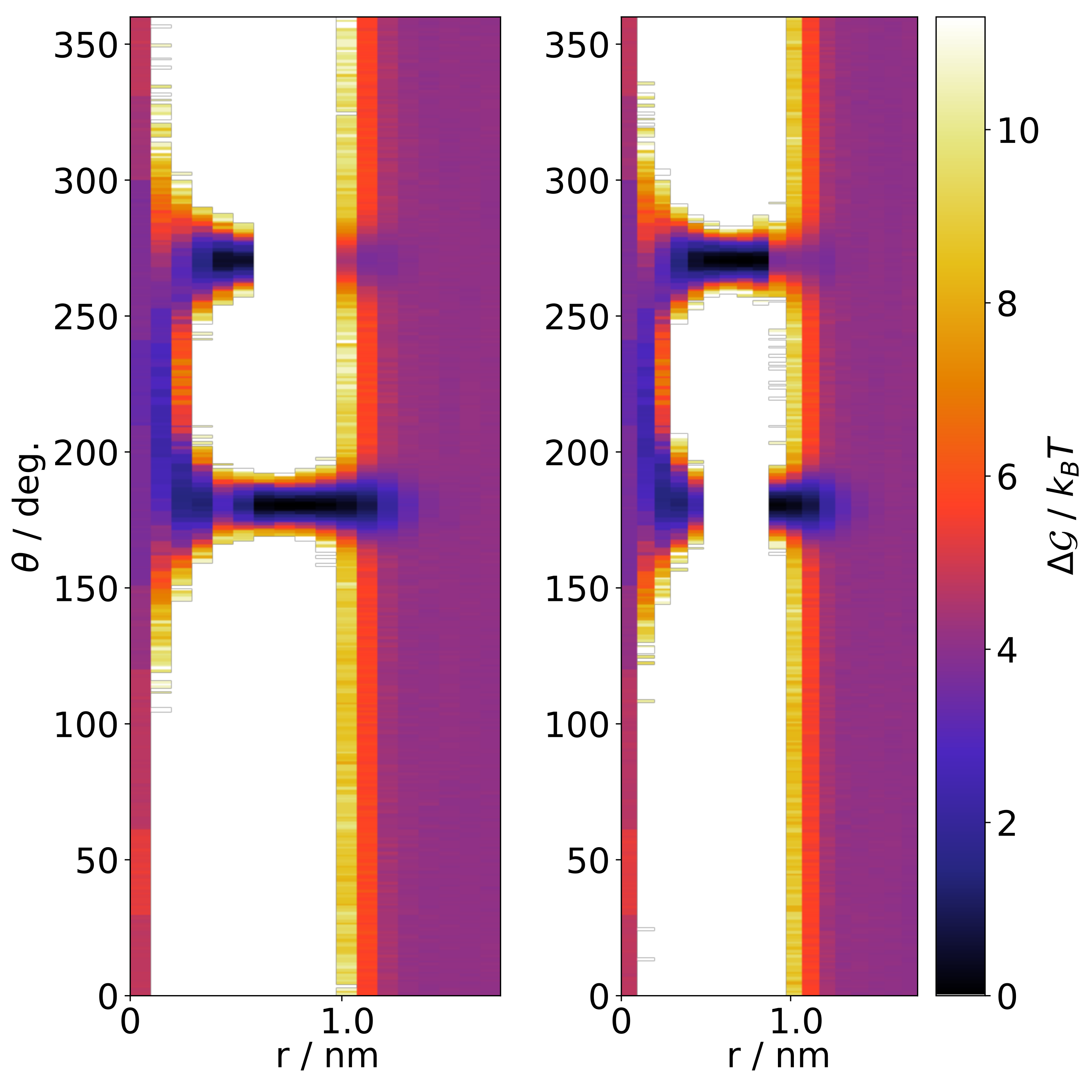}
		\caption{\baselineskip4mm Thermodynamic integration
                  study of the 2D model of ligand-protein
                  dissociation. (Left) Free energy profiles obtained
                  for various TI protocols: In ``ideal TI'', we
                  sampled over all angles for each $r$. Apart from
                  numerical issues, this result should therefore agree
                  with the exact result obtained from the partition
                  function. Moreover, two Monte Carlo TI calculations
                  are shown, which were started at path 1 and 2,
                  respectively. Since TI is in principle exact, these
                  curve should also coincide with the exact
                  result. However, both calculations fail, due to
                  jumps between the two path. (Right) 2D
                  representation of the energy landscape of the model,
                  reflecting the jumps of the Monte Carlo TI
                  calculations starting at path 1 ($\sim 180^\circ$)
                  and 2 ($\sim 270^\circ$). Note that for each value
                  of $r$, the Monte Carlo moves can only go
                  vertically. }
		\label{fig:2DTI}
	\end{figure}


	\subsection{Rate estimation}
	
	To compare the kinetics of the 2D model defined
        by the potential $V(r_1,r_2)$ [Eq.~\eqref{eq:2DHam}], and the
        path-separated 1D models using the respective pathway
        free energies $\Delta \mathcal{G}_k(r)$, we performed standard
        Langevin simulations governed by 
	\begin{align}
		\text{2D:} \quad m \ddot{\vec{r}} &= -\frac{\partial V}{\partial \vec{r}} - \itGamma \dot{\vec{r}} + \vec{\eta} \; , \\
		\text{path 1:} \quad m \ddot{r} &= -\frac{\partial \mathcal{G}_1}{\partial r} - \itGamma r + \eta \; , \quad\quad\text{path 2:} \quad m \ddot{r} = -\frac{\partial \mathcal{G}_2}{\partial r} - \itGamma r + \eta \; ,
	\end{align}
	where $\eta = \sqrt{2 \kb T \, \itGamma}\mathcal{R}$. We used a mass
        $m=0.8$~kg/mol, constant friction $\itGamma_k =
        $~20~kg/(mol\,ps) and temperature $T=300$~K. 
	
	
        In the 2D case, we used a simple Euler integration scheme to
        propagate $10 \times 1\,\mu$s-long trajectories, where 42
        start in the associated and 8 in the dissociated state,
        employing an integration time-step of 1~fs. We saved the
        positions $(r_1,r_2)$ for every ns. At $r=1.75$~nm, we introduced
        a reflective wall by a flip of the position and velocity
        components perpendicular to the wall.
	We calculated the waiting times of the associated and
        dissociated state, by tracking the number of frames in a state
        until leaving, using geometrical cores at
        $r=0.2$~nm and 1.4~nm, respectively. The rates
        $k_\text{off}$ and
        $k_\text{on}$ are then estimated by the inverse of the average
        waiting time, for each trajectory individually. We report
        their average and standard deviation in Table~I of the main
        text.

	
	We also performed biased simulations of the 2D system by constraining the distance to the center as $r=r_0+vt$, where we use $r_0=0.01$~nm and $v=0.01$~m/s. In particular, we integrate the constrained Langevin dynamics by transforming the equations of motion to polar coordinates, calculating the forces acting on $r$ and $\theta$ as usual, but then updating $r$ to conform to its constraint and recording the forces on $r$ as the constraint force. We finally compute the work by integrating these forces. Sampling from an initial unbiased distribution in $\theta$ and $\dot{\theta}$ at fixed $r_0$, we pull 10000 trajectories towards $r=1.75$~nm. By separating the trajectories using pathway 1 and 2, respectively, we use Jarzynski's equality to calculate the pathway free energy.

	
	Using these estimates of the pathway free energy, we performed
        one-dimensional Langevin simulations using the mass
        and friction given above. We simulated $10\times 1$~ms long simulations
        for both pathway 1 (45 starting in the associated, 5 in the
        dissociated state) and pathway 2 (40, 10), again with $\delta t =
        1$~fs and a reflective wall at $1.75$~nm. We computed the rates
        and weigh them by the probabilities obtained by Eq.~(26) of the
        main text, reported in Table~I of the main text.


	\begin{figure}[htb!]
		\centering
		\includegraphics[width=0.5\textwidth]{\dirfig/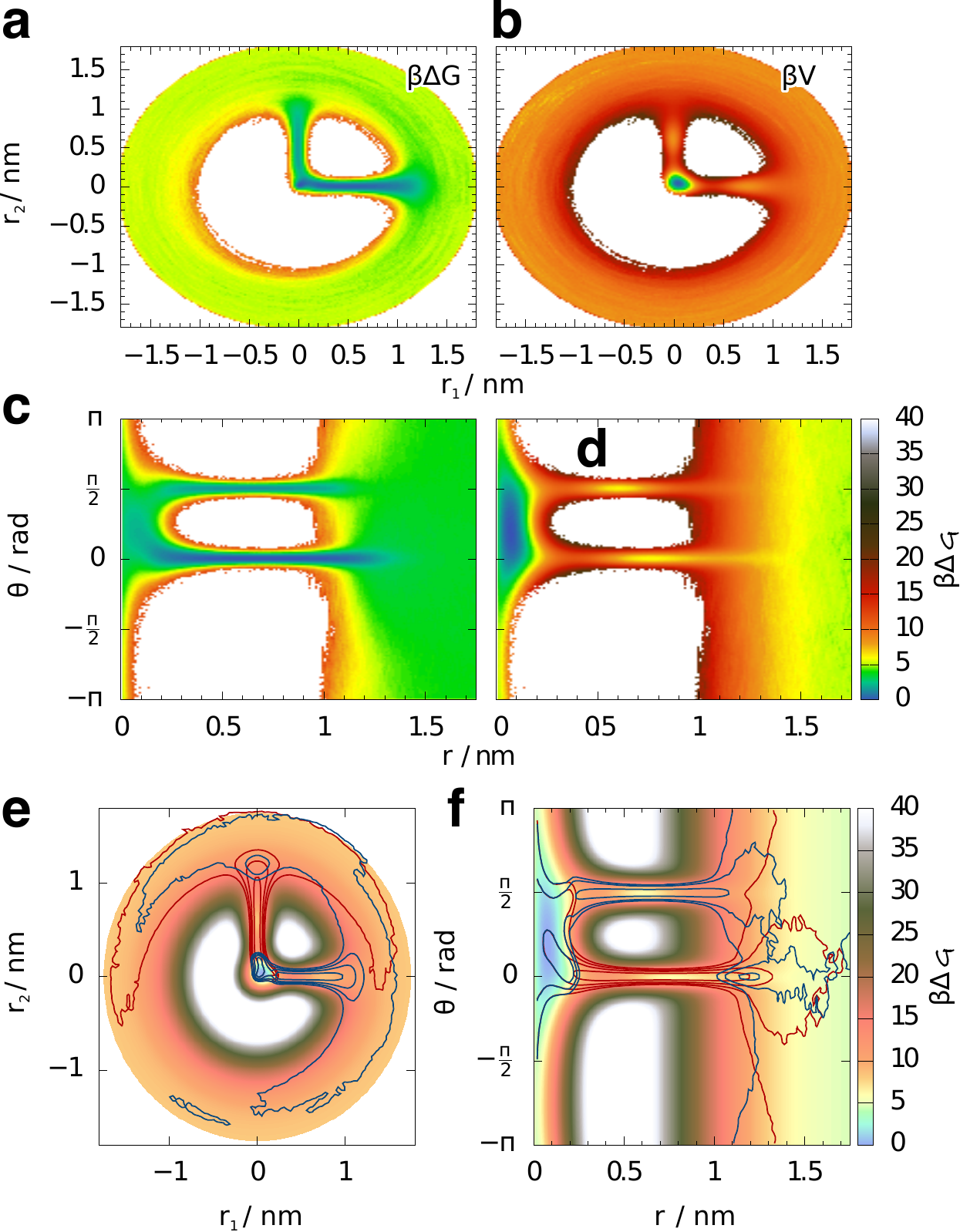}
		\caption{\baselineskip4mm Sampling of the 2D potential energy landscape in nonequilibrium pulling. (a) Distribution in $(r_1,r_2)$ of all pulled trajectories as a biased density $\mathcal{V}$, estimated by a histogram. (b) Corresponding reweighed potential $V$ using Eq.~(11) of Ref.~\onlinecite{Post19}. (c) Same data but projected onto $(r,\theta)$ coordinates. (d) Corresponding reweighed density. (e) Biased trajectories separated by pathways superimposed on unbiased potential in the $(r_1,r_2)$ plane, visualized by isolines of same biased density. (f) Same data but in $(r,\theta)$.}
		\label{fig:2d_pulling}
	\end{figure}
	
\clearpage
%
%
\section{Normality test of the work distribution $P(W)$}
	
In addition to estimating the probability distribution $P(W)$ via a
histogram of the sampled work $W$ at a specific value of coordinate
$x$, we can evaluate the normal probability plot of that sample. To
this end, we employ the \url{scipy.stats} function \url{probplot},
which compares the points of the computed work to the quantiles of the
corresponding normal distribution (the probit), such that the
deviations reveal a non-Gaussian shape. In particular, we sort the
work sample in increasing order and plot them against equidistant
points of the inverse of the cumulative distribution
function. Deviations from a straight line reveal an underlying
non-Gaussian distribution.

\begin{figure}[htb!]
		\centering
		\includegraphics[width=0.5\textwidth]{\dirfig/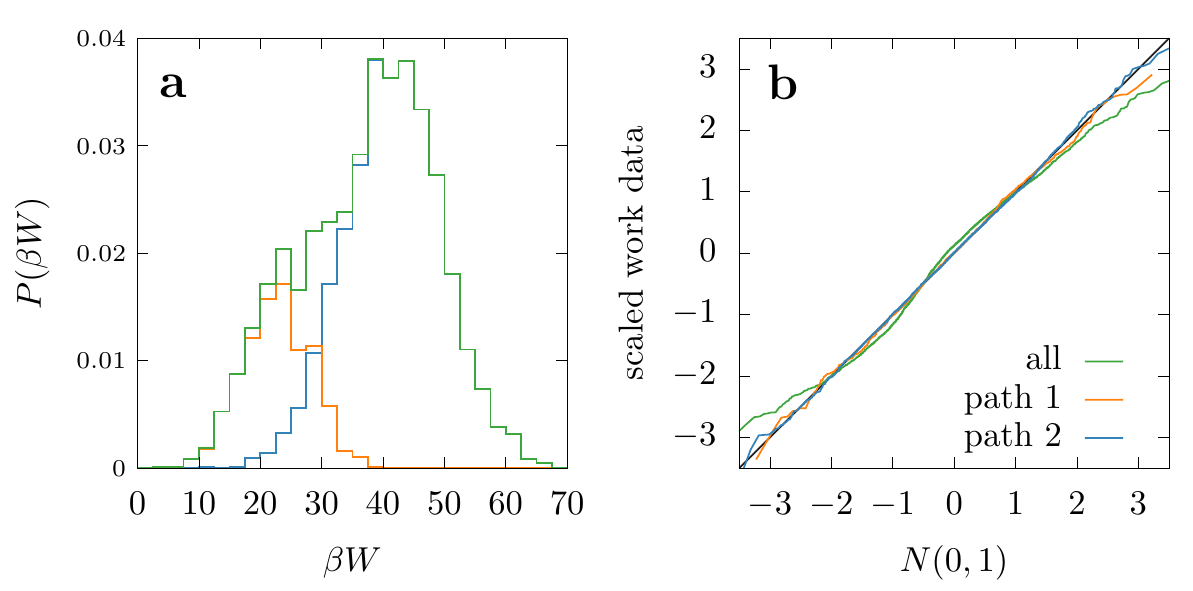}
		\includegraphics[width=0.5\textwidth]{\dirfig/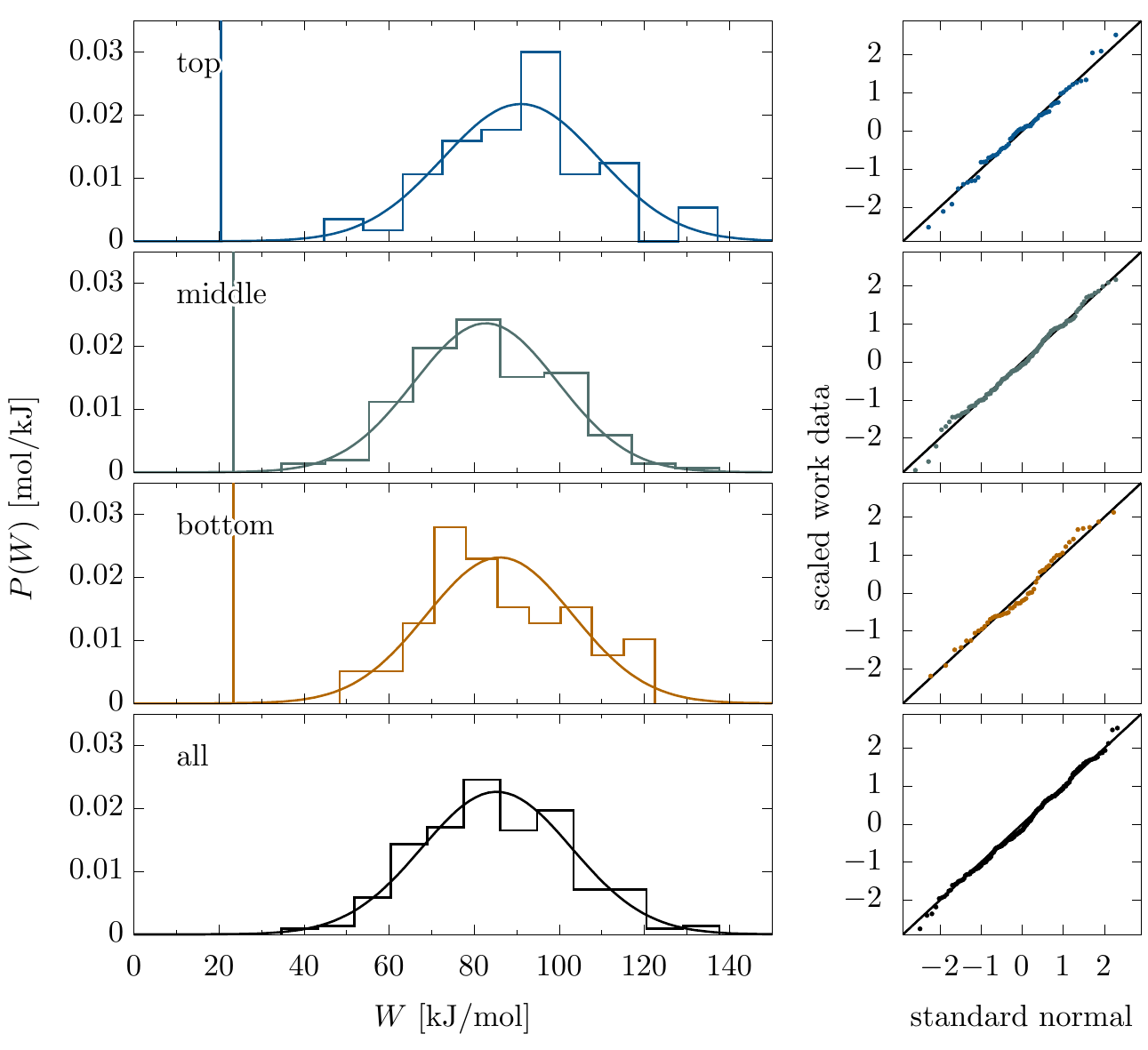}
                \caption{\baselineskip4mm Normality test of $P(W)$
                  obtained for (top panels) the 2D model and (bottom
                  panels) for trypsin. (a) $P(W)$ for complete and
                  path-separated trajectory sets. (b) Normality tests
                  for these sets. The complete trajectory set displays
                  clear deviations in the tails from the ideal
                  diagonal, while the path-separated sets display
                  better agreement.  (Bottom panels) Results for
                  trypsin, showing (left) $P(W)$ for complete and
                  path-separated trajectory sets and (right) normality
                  tests for these sets. Both the path-separated the
                  complete trajectory sets display rather good
                  agreement with a normal distribution.}
		\label{fig:2d_normtest}
\end{figure}

\clearpage        
%
%
\section{Trypsin-benzamidine complex:\\ PCA on ligand-protein contacts}

	\begin{figure}[htb!]
		\centering
		\includegraphics[width=0.45\textwidth]{\dirfig/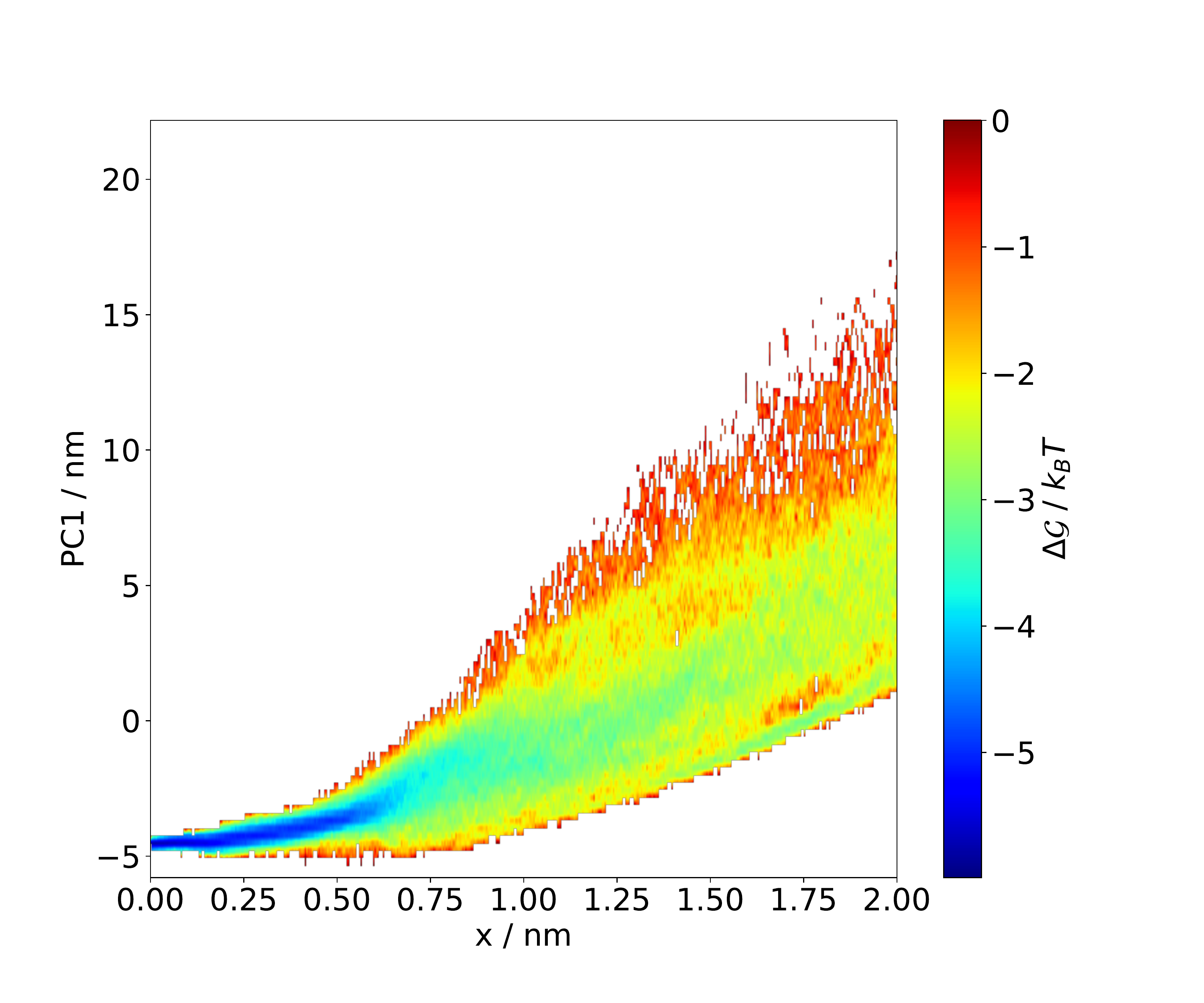}
		\includegraphics[width=0.38\textwidth]{\dirfig/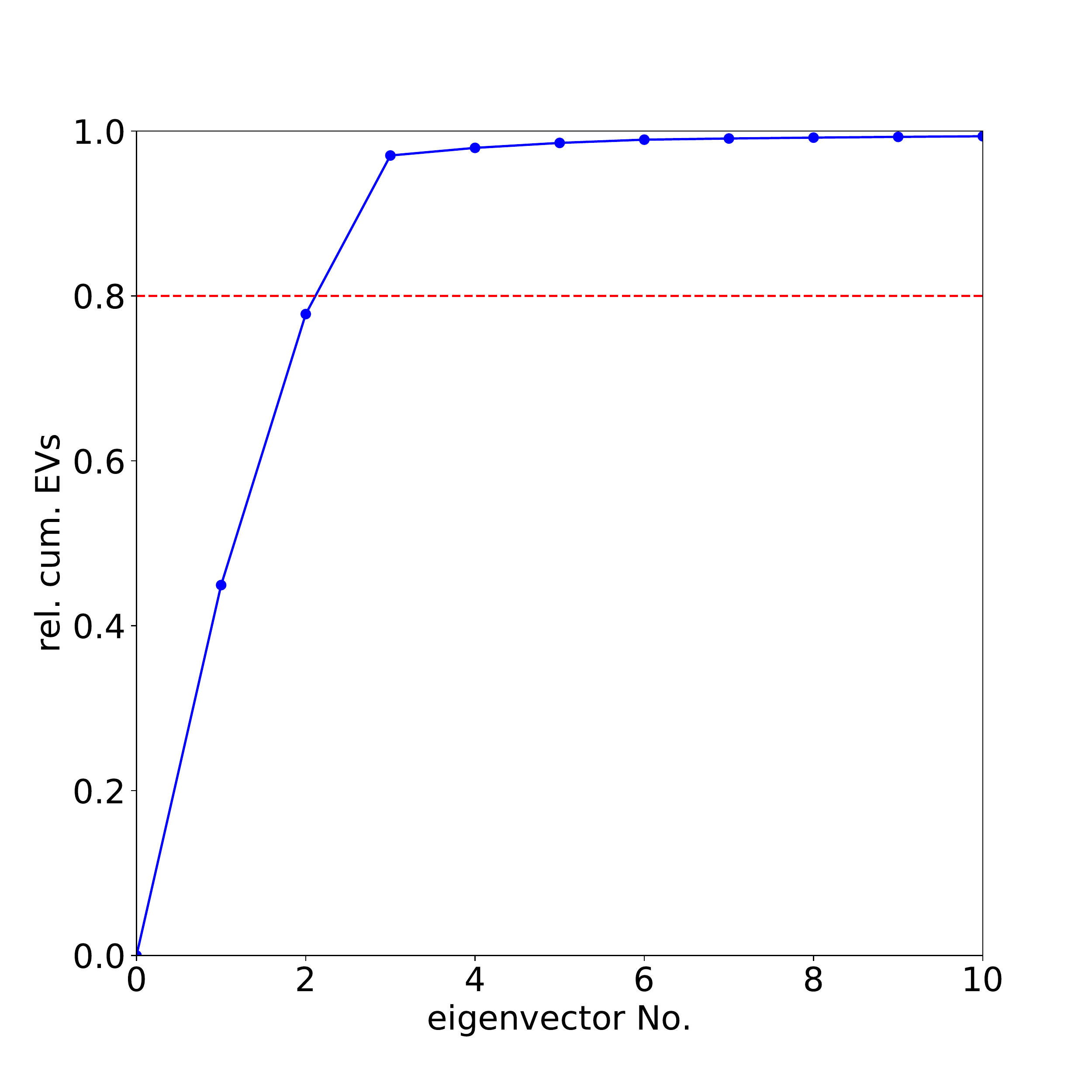}
		\includegraphics[width=0.85\textwidth]{\dirfig/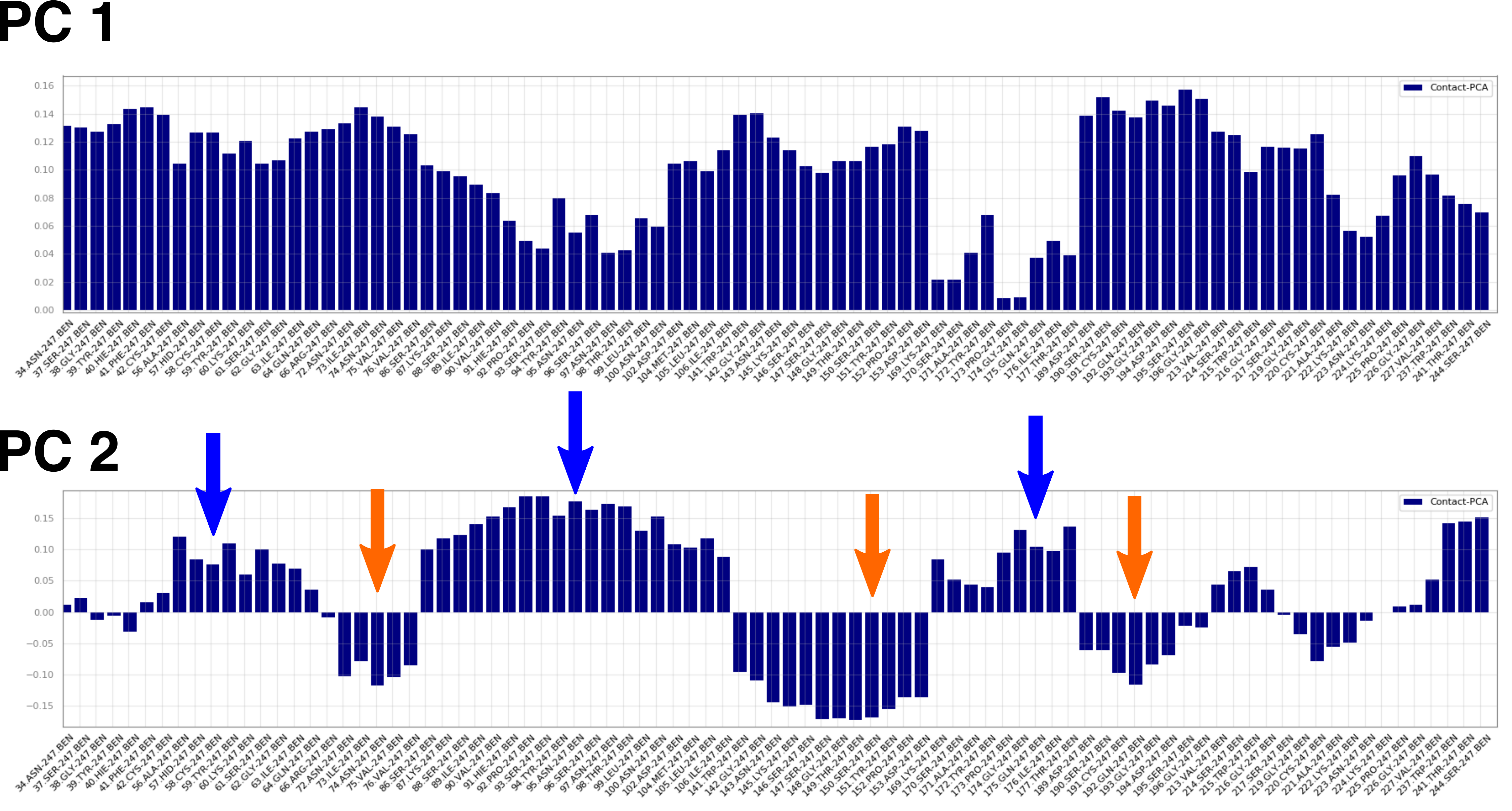}
		\caption{\baselineskip4mm Results from a contact PCA,
                  obtained for the trypsin-benzamidine complex. (Top
                  left) Correlation between pulling distance $x$ and
                  1\up{st} principal component. (Top right) Cumulative
                  fluctuations obtained from the principal
                  components. The first two components already explain
                  78\% of the variance of the data. (Bottom)
                  Eigenvector contributions from single
                  contacts. Residues defining pathways displayed in
                  Fig.~2 of the main text are indicated by accordingly
                  colored arrows.}
		\label{fig:TrypPC1x}
	\end{figure}

\newpage

\section*{Supporting References}

\end{document}